\documentclass[sigconf,natbib=true,anonymous=false]{acmart}
\AtBeginDocument{%
  \providecommand\BibTeX{{%
    \normalfont B\kern-0.5em{\scshape i\kern-0.25em b}\kern-0.8em\TeX}}}

\setcopyright{acmcopyright}
\copyrightyear{2018}
\acmYear{2018}
\acmDOI{XXXXXXX.XXXXXXX}

\acmConference[Conference acronym 'XX]{Make sure to enter the correct
  conference title from your rights confirmation emai}{June 03--05,
  2018}{Woodstock, NY}
\acmPrice{15.00}
\acmISBN{978-1-4503-XXXX-X/18/06}

\usepackage{svg}
\usepackage{subfigure}
\usepackage{bbding}
\usepackage{enumitem}
\usepackage{multirow}
\usepackage{makecell}
\usepackage{url}            
\usepackage{booktabs}       
\usepackage{amsfonts}       

\title{RL4RS: A Real-World Dataset for Reinforcement \\ Learning based Recommender System}

\author{Kai Wang$^1$, Zhene Zou$^1$, Minghao Zhao$^1$, Qilin Deng$^1$, Yue Shang$^1$, Yile Liang$^1$, \\ Runze Wu$^1$, Xudong Shen$^1$, Tangjie Lyu$^1$, Changjie Fan$^1$}

\affiliation{%
    \institution{$^1$Fuxi AI Lab, NetEase Inc., Hangzhou, China}
    \country{}
  \institution{\{wangkai02,zouzhene, zhaominghao,dengqilin,shangyue,liangyile, \\ wurunze1,hzshenxudong	,hzlvtangjie,fanchangjie\}}@corp.netease.com  
}

\renewcommand{\shortauthors}{Wang, et al.}

\begin{document}

\begin{abstract}
Reinforcement learning based recommender systems (RL-based RS) aim at learning a good policy from a batch of collected data, by casting recommendations to multi-step decision-making tasks. However, current RL-based RS research commonly has a large reality gap. 
In this paper, we introduce the first open-source real-world dataset, RL4RS, hoping to replace the artificial datasets and semi-simulated RS datasets previous studies used due to the resource limitation of the RL-based RS domain. 
Unlike academic RL research, RL-based RS suffers from the difficulties of being well-validated before deployment. We attempt to propose a new systematic evaluation framework, including evaluation of environment simulation, evaluation on environments, counterfactual policy evaluation, and evaluation on environments built from test set.
In summary, the RL4RS (Reinforcement Learning for Recommender Systems), a new resource with special concerns on the reality gaps, contains two real-world datasets, data understanding tools, tuned simulation environments, related advanced RL baselines, batch RL baselines, and counterfactual policy evaluation algorithms. The RL4RS suite can be found at \textit{\url{https://github.com/fuxiAIlab/RL4RS}}. 
In addition to the RL-based recommender systems, we expect the resource to contribute to research in applied reinforcement learning.
\end{abstract}




\maketitle

\begin{figure}[h]
  \centering
  \includegraphics[scale=0.25]{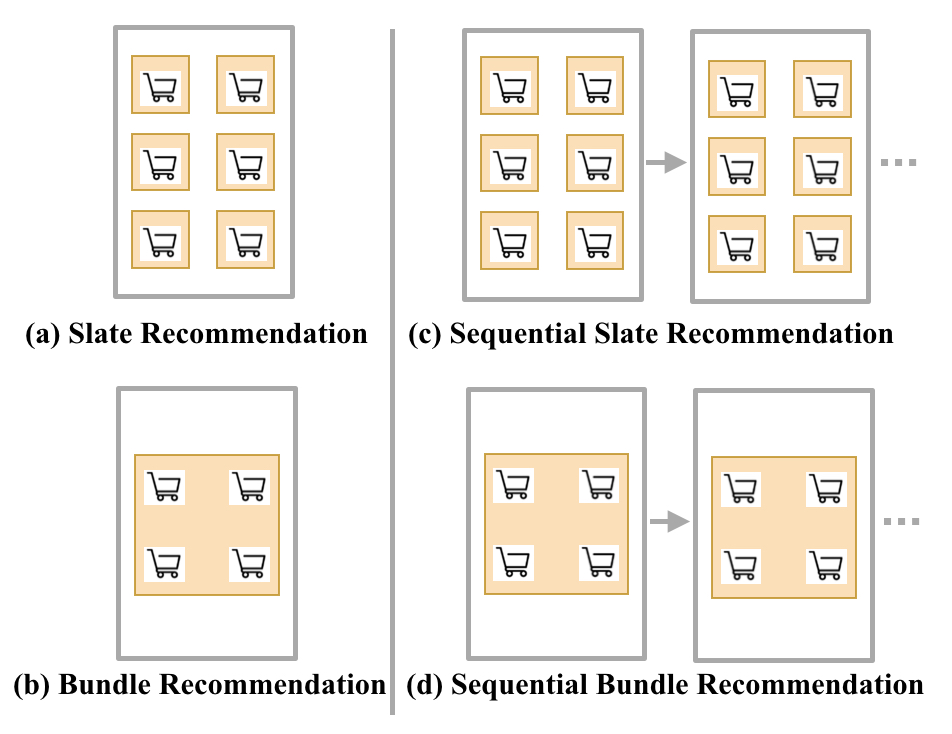}
  \caption{Several novel recommendation scenarios. \label{Fig:scenes}}
\end{figure}

\begin{table*}[]
\centering
\caption{A comparison between RL4RS and other resources.}
\label{tab:impact}
\begin{tabular}{lccccccccl} \toprule
                 & \multicolumn{3}{c}{Dataset}                                                                                                                                                             & \multicolumn{3}{c}{Opensource}                                                                                                     & \multicolumn{2}{c}{Others}                                                                                                             \\ \cmidrule(lr){2-4}\cmidrule(lr){5-7}\cmidrule(lr){8-9} 
               Resource  & \begin{tabular}[c]{@{}c@{}}Artifical\\ dataset\end{tabular} & \begin{tabular}[c]{@{}c@{}}Semi-simulated\\ dataset\end{tabular} & \begin{tabular}[c]{@{}c@{}}Real \\ dataset\end{tabular} & Code & \begin{tabular}[c]{@{}c@{}}Raw logged\\ data\end{tabular} & \begin{tabular}[c]{@{}c@{}}Simulation\\ environment\end{tabular} & \begin{tabular}[c]{@{}c@{}}Offline policy\\ learning\end{tabular} & \begin{tabular}[c]{@{}c@{}}Offline policy\\ evaluation\end{tabular} \\ \midrule
RecoGym\cite{Rohde2018RecoGymAR}          & \checkmark                                                           & $\times$                                                           & $\times$                                                        & \checkmark    & $\times$                                                         & \checkmark                                                                & $\times$                                                                 & $\times$                                                                   \\ 
Recsim\cite{ie2019recsim}           & \checkmark                                                           & $\times$                                                                & $\times$                                                       & \checkmark    & $\times$                                                          & \checkmark                                                                & $\times$                                                                  & $\times$                                                                    \\ 
Virtual-Taobao\cite{Shi2019VirtualTaobaoVR}   & $\times$                                                           & $\times$                                                                & \checkmark                                                       & \checkmark    & $\times$                                                         & \checkmark                                                                & $\times$                                                                 & $\times$                                                                   \\ 
Top-k Off-policy\cite{chen2019topk} & \checkmark                                                           & $\times$                                                                & $\times$                                                        & \checkmark    & $\times$                                                         & $\times$                                                                & \checkmark                                                                 & $\times$                                                                   \\ 
SlateQ\cite{Ie2019SlateQAT}          & \checkmark                                                           & $\times$                                                                & $\times$                                                       & \checkmark    & $\times$                                                         & $\times$                                                                & $\times$                                                                 & $\times$                                                                   \\ 
Adversarial Model\cite{chen2019adversarialgenerative}      & $\times$                                                           & \checkmark                                                                & $\times$                                                        & \checkmark    & $\times$                                                         & $\times$                                                                 & $\times$                                                                 & $\times$                                                                   \\ 
List-wise\cite{zhao2017list}        & $\times$                                                           & \checkmark                                                                 & $\times$                                                       & \checkmark    & $\times$                                                         & $\times$                                                                & $\times$                                                                 & $\times$                                                                   \\ 
Model-based RS\cite{Bai2019ModelBasedRLnips}      & \checkmark                                                           & \checkmark                                                                 & $\times$                                                       & \checkmark    & $\times$                                                         & $\times$                                                                & \checkmark                                                                  & $\times$                                                                     \\ \hline
Ours             & $\times$                                                           & $\times$                                                                 & \checkmark                                                       & \checkmark    & \checkmark                                                         & \checkmark                                                                & \checkmark                                                                 & \checkmark                                                                   \\ \bottomrule
\end{tabular}
\end{table*}

\section{Introduction}
\label{sec1}

In 2022, retail e-commerce sales worldwide amounted to 5.7 trillion US dollars, and e-retail revenues are projected to grow to 6.3 trillion US dollars in 2023. Such rapid growth promises an excellent future for the worldwide e-commerce industry, signifying a strong market and increased customer demand. Besides the massive increment of traffic volume, there has been a rapid growth of various recommendation scenarios, including slate recommendation (lists of items), bundle recommendation (a collection of items that should be purchased simultaneously), sequential recommendation, and many others, as shown in Figure \ref{Fig:scenes}. It is worth exploring the various challenges the modern e-commerce industry faces today.
Most current e-commerce and retail companies build their recommender systems by implementing supervised learning based algorithms on their websites to maximize immediate user satisfaction in a greedy manner. However, the item-wise greedy recommendation strategy is an imperfect fitting to real recommendation systems. With more and more new upcoming recommendation scenarios, more and more challenges have to be solved. For instance, in sequential recommendation scenarios, traditional methods often consider different recommendation steps in a session to be independent and fail to maximize the expected accumulative utilities in a recommendation session. In the slate recommendation or bundle recommendation scenarios, the conversion rate of an item does not solely depend on itself. If an item is surrounded by similar but expensive items, the conversion rate increases, known as the decoy effect~\cite{2020Validation}. However, the possible combinations of all the items can be billions, which is an NP-hard problem and less explored in supervised learning (SL). 

\begin{figure}[h]
  \centering
  \includegraphics[scale=0.4]{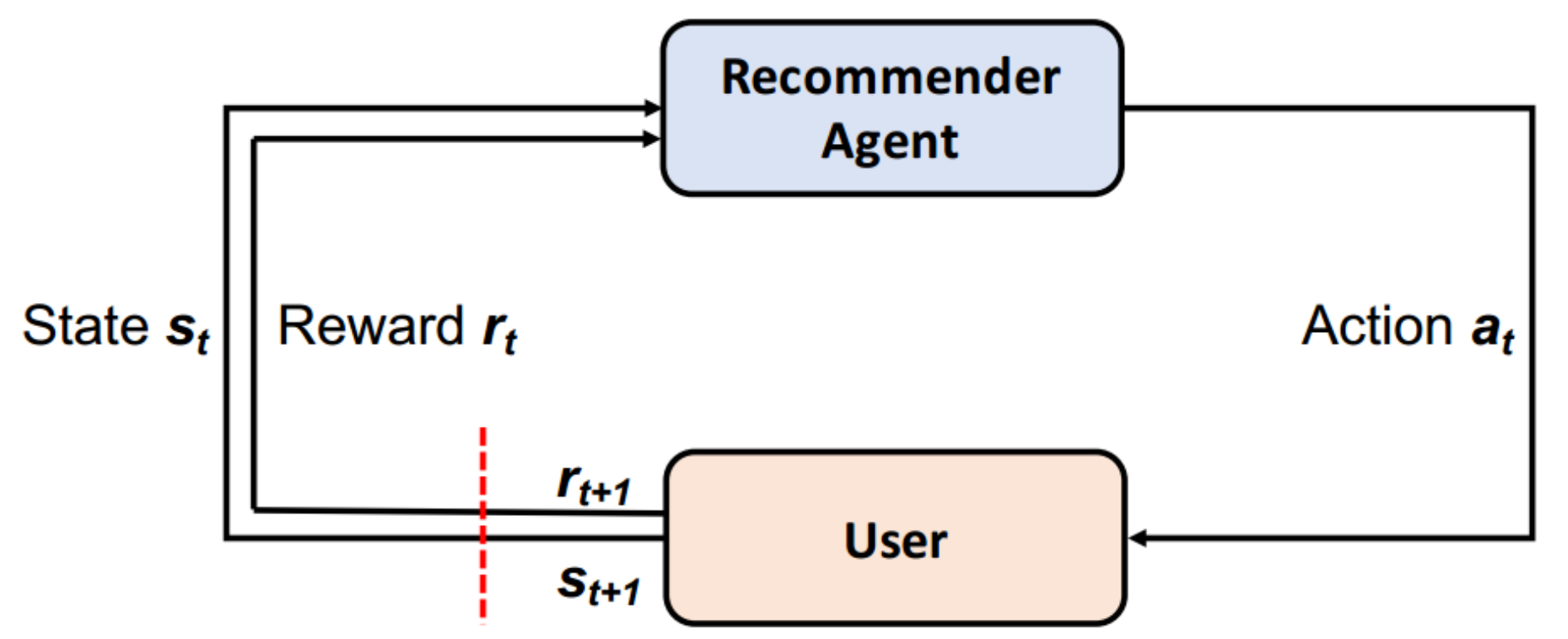}
  \caption{The user-agent interaction in the Markov Decision Process (MDP) of RL-based recommender systems. \label{Fig:rl-based-rs}}
\end{figure}

To deal with these challenges, recent researchers resort to adopting reinforcement learning for recommendations, in which the recommendation process is formulated as a sequential interaction between the user (environment) and the recommendation agent (RL agent), as illustrated in Figure \ref{Fig:rl-based-rs}.
Reinforcement learning is a promising direction since the RL paradigm is inherently suitable for tackling multi-step decision-making problems, optimizing long-term user satisfaction directly, and exploring the combination spaces efficiently 
 - but there remain two problems in recent research.

The first problem is the lack of real-world datasets for RL-based RS problems. There are mainly two alternatives, one is artificial datasets, such as RecoGym~\cite{Rohde2018RecoGymAR} and RECSIM~\cite{ie2019recsim}. The main disadvantage is that they are not the real feedback of users in real applications. 
Another is semi-simulated datasets, i.e., traditional RS datasets (e.g., MovieLens) that transformed to RL data format. Take MovieLens dataset as an example, to meet the requirements of RL data format, The Adversarial User Model~\cite{chen2019adversarialgenerative} introduces external movie information and assumes the context of user's choice as the movies released within a month. The maximal size of each displayed set is set as 40. The main disadvantage of semi-simulated datasets is that many forced data transformations are unreasonable.

The second problem is the lack of unbiased evaluation methods. In the current research, there are mainly two kinds of evaluation indicators: traditional recommendation indicators (Recall Rate, Accuracy, etc.) and pure reinforcement learning indicators (e.g., Cumulative Rewards). However, the former are short-term evaluation indicators, and the latter highly depend on the accuracy of the simulation environment. The bias of policy evaluation also comes from "extrapolation error", a phenomenon in which unseen state-action pairs are erroneously estimated to have unrealistic values. In this paper, we propose a new evaluation framework and explore two other recently developed methods to tackle "extrapolation error", counterfactual policy evaluation and batch RL.

With these in mind, we introduce RL4RS - an open-source dataset for RL-based RS developed and deployed at Netease. RL4RS is built in Python and uses TensorFlow for modeling and training. It aims to fill the rapidly-growing need for RL systems that are tailored to work on novel recommendation scenarios. It consists of (1) two large-scale raw logged data, reproducible simulation environments, and related RL baselines. 
(2) data understanding tools for testing the proper use of RL, and a systematic evaluation framework, including evaluation of environment simulation, policy evaluation on simulation environments, and counterfactual policy evaluation.
(3) the separated data before and after reinforcement learning deployment for each dataset, e.g., Slate-SL and Slate-RL. Based on them, we are able to evaluate the effectiveness of different batch RL algorithms and measure the extent of extrapolation error, by taking Slate-SL as train set and Slate-RL as test set. 

\section{Impact}
\label{sec2}

In this section, we 
assess the benefits of the RL4RS resource in relation to existing resources for evaluating RL-based RS. We collect all relevant works that have been open-sourced at present, including 
dataset/environment resources 
(RecoGym\renewcommand{\thefootnote}{$1$}\footnote{https://github.com/criteo-research/reco-gym}\cite{Rohde2018RecoGymAR}, Recsim\renewcommand{\thefootnote}{$2$}\footnote{https://github.com/google-research/recsim}\cite{ie2019recsim}, and Virtual Taobao\renewcommand{\thefootnote}{$3$}\footnote{https://github.com/eyounx/VirtualTaobao}\cite{Shi2019VirtualTaobaoVR}) 
and representative method/algorithm resources 
(Top-k off-policy\renewcommand{\thefootnote}{$4$}\footnote{https://github.com/awarebayes/RecNN}\cite{chen2019topk}, 
SlateQ\renewcommand{\thefootnote}{$5$}\footnote{https://github.com/google-research/recsim}\cite{Ie2019SlateQAT},
Adversarial User Model\renewcommand{\thefootnote}{$6$}\footnote{https://github.com/xinshi-chen/GenerativeAdversarialUserModel}\cite{chen2019adversarialgenerative}, 
List-wise\renewcommand{\thefootnote}{$7$}\footnote{https://github.com/luozachary/drl-rec}\cite{zhao2017list}, 
and 
Model-Based RS\renewcommand{\thefootnote}{$8$}\footnote{https://github.com/XueyingBai/Model-Based-Reinforcement-Learning-for-Online-Recommendation}\cite{Bai2019ModelBasedRLnips}).
For method/algorithm resources, we mainly focus on the datasets used in the offline experiments and how they are used.

We consider the following dimensions to evaluate the benefit of these related works:
\begin{enumerate}[leftmargin=*]
    \item \textbf{Artificial Datasets}: RecoGym and Recsim are two representative artificial datasets which are employed in the experiments of Top-K and SlateQ.
    
    \item \textbf{Semi-simulated Datasets}: Traditional RS datasets, such as MovieLens, are designed for item-wise supervised learning and are not suitable for RL-based RS experiments. As an alternative solution, Adversarial User Model, List-wise, and Model-based RS make a lot of assumptions and manual transformations on these datasets to fit the requirements of RL.
    
    \item \textbf{Real Industry Datasets Without Transformation}: Though many works build their online experiments on real industrial scenarios, little research provides the raw data for reproducing offline experiment results.
    
    \item \textbf{Code Release}: We list the GitHub pages of each algorithm at the beginning of this section, where Top-k Off-policy is a non-official implementation.
    
    \item \textbf{Dataset Release}: In addition to artificial datasets and semi-simulated RS datasets, Virtual Taobao builds experiments on a real dataset but without open-sourcing the raw logged data. Raw logged data is necessary for offline policy learning and evaluation, and the reproduction of the simulation environments.
    
    \item \textbf{Simulation Environment Release}: Virtual Taobao has open-sourced a low-dimensional pre-trained simulation environment, which is associated with 11-dimensional user features and 27-dimensional item features.
    
    \item \textbf{Offline Policy Learning}: Most current works first train the environmental model and then learn the policy through the interaction with the pre-trained environment model, except Top-k Off-policy. In this paper, we further explore batch RL.
    
    \item \textbf{Offline Policy Evaluation}: Offline policy evaluation aims to predict the performance of a newly learned policy without having to deploy it online. Rather than test the policy in a simulation environment, in this paper, we further introduce the counterfactual policy evaluation (CPE) to RL-based RS.
\end{enumerate}

In Table 1, we summarize the characteristics of existing works in terms of these dimensions. It can be seen that our RL4RS resource is the only one that meets all requirements (especially real-world dataset release). In addition to the contribution of open-sourcing two industrial datasets, we further explore the topics such as offline policy training and counterfactual policy evaluation, hoping to enhance the development of RL-based RS field.

\begin{table*}[tb]
\centering
\caption{RL4RS - Dataset Details}
\label{tab:dataset}
\begin{tabular}{lcccccc}
\toprule
Num. of & Slate & Slate-SL & Slate-RL & SeqSlate & SeqSlate-SL & SeqSlate-RL \\ \hline
Users & 149,414 & 112,221 & 77,834 & 149,414 & 112,221 & 77,834 \\
Items & 283 & 283 & 283 & 283 & 283 & 283 \\
Vaild item slates & \textasciitilde $94^9$ & \textasciitilde $94^9$ & \textasciitilde $94^9$ & \textasciitilde $94^{36}$ & \textasciitilde $94^{36}$ & \textasciitilde $94^{36}$ \\
Sessions & 1719,316 & 937,949 & 781,367 & 958,566 & 519,435 & 439,131 \\
Items per session & 9.0 & 9.0 & 9.0 & 16.1 & 16.3 & 16.0 \\
Purchases per session & 5.46 & 4.89 & 6.09 & 9.75 & 8.83 & 10.84 \\
Rewards per session & 91.7 & 83.2 & 102.0 & 164.5 & 150.3 & 181.5 \\
\bottomrule
\end{tabular}
\label{table:mlm_datasets}
\vspace{-4mm}
\end{table*}

\section{Data Description}
\label{sec3}
We collect the raw logged data from
one of the most popular games released by \textit{NetEase Games}\renewcommand{\thefootnote}{$9$}\footnote{http://leihuo.163.com/en/index.html}.
To ensure the privacy of users, we adopt a three-phase anonymization procedure including user sampling, value encoding, and feature description masking. And all users are provided with the option to turn off data collection.
We employ a non-reversing encoding algorithm, MD5 \cite{rivest1992md5} and logarithm function, to encode the original discrete and continuous values, respectively.
The item recommendation task in this scenario is characterized by its special interaction rules.
In each second, the recommendation engine should respond to users' requests with 3 item lists (3 items per list), and the next item list is locked until the items of the current list are sold out, which we refer to as the "unlock" rule.
Obviously, due to the special rule, the users' response to an item depends on not only that item but also the items of other lists.
If users are unsatisfied with the existing item slate, they can refresh the page (i.e., the item slate) through a refresh button up to 3 times per day.

\begin{figure}[tb]
  \centering
  \includegraphics[scale=0.22]{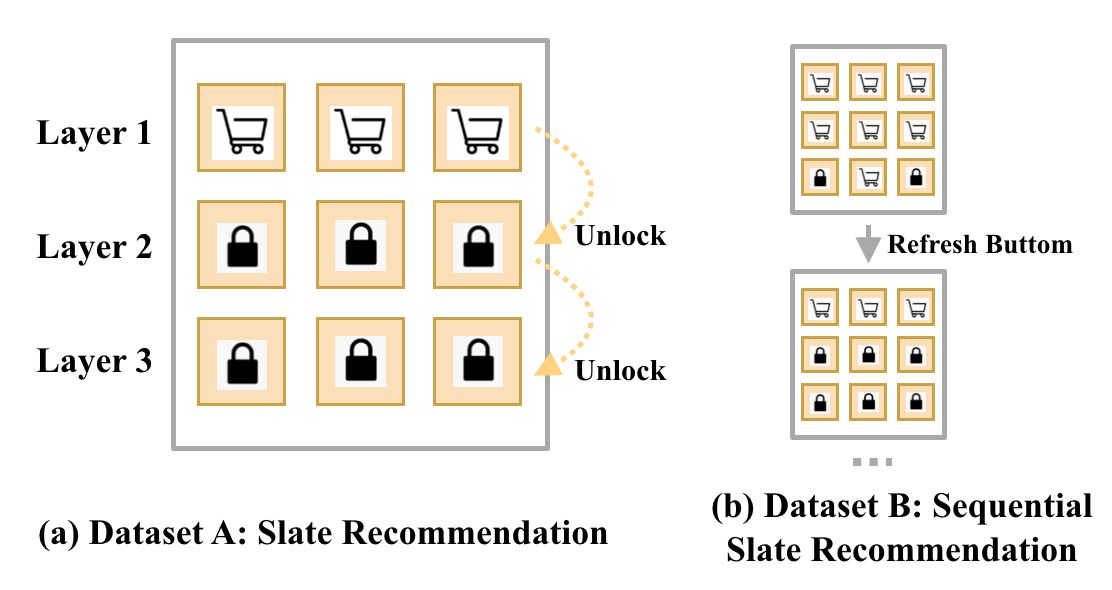}
  \caption{A graphic description of RL4RS datasets. "Unlock" means the next layer becomes available only when all the items in the current layer have been sold out.\label{Fig:dataset}}
\end{figure}

Here, we provide two datasets, RL4RS-Slate and RL4RS-SeqSlate, as shown in Figure \ref{Fig:dataset}. For each dataset, the separated data before and
after RL deployment are also available.  RL4RS-Slate focuses on the slate recommendation. It regards the user's behavior on a single page as an MDP process. RL4RS-SeqSlate focuses on the sequential slate recommendation. It not only considers how to recommend a single page but also considers the relationship between pages to maximize the total reward of the session.
A brief statistics of datasets are shown in Table \ref{tab:dataset}.


\subsection{Logged Data}
\begin{figure}[h]
  \centering
  \includegraphics[scale=0.45]{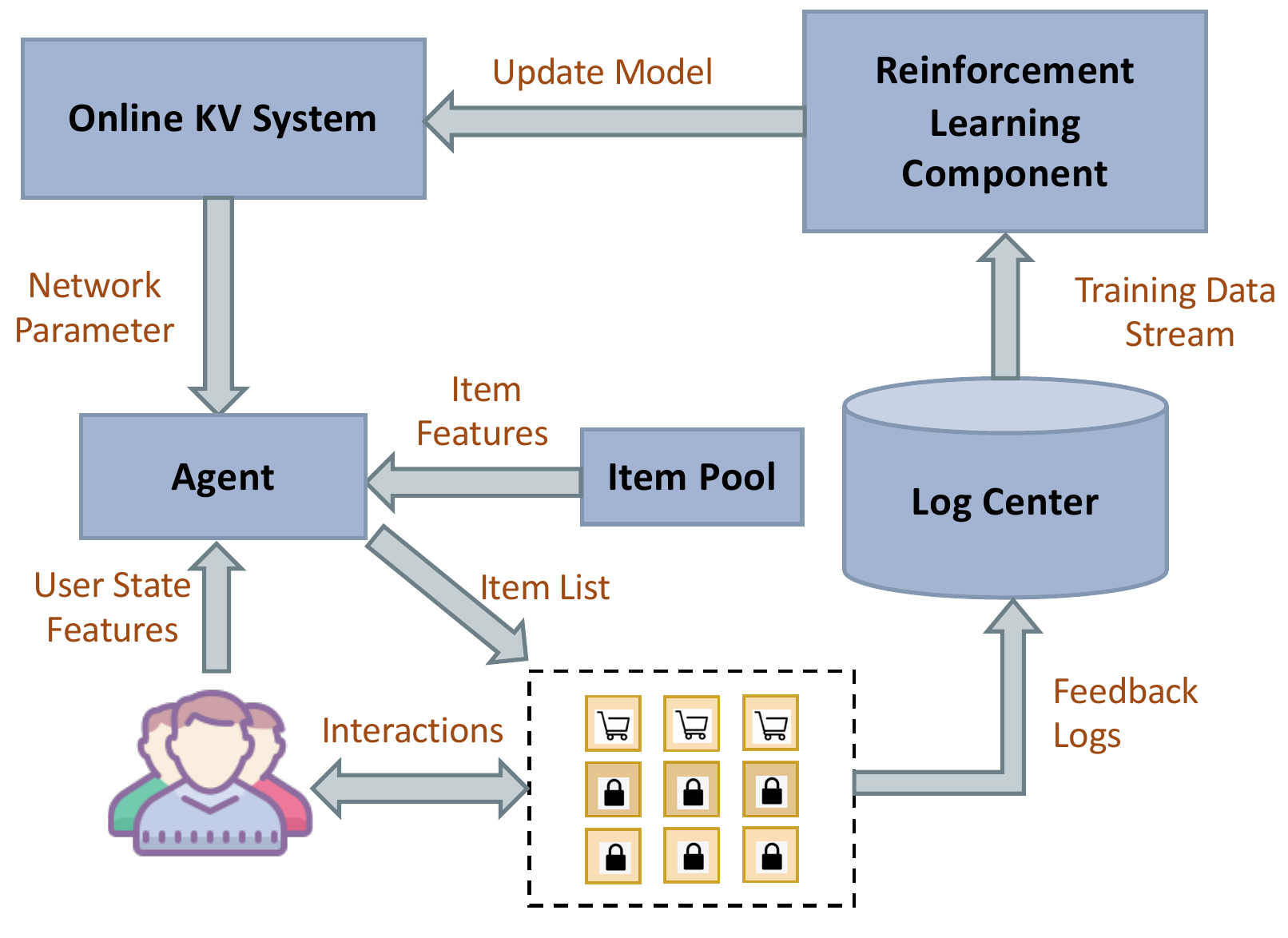}
  \caption{The architecture of recommender system for collecting data. \label{Fig:RL_Ranking_System}}
\end{figure}

Roughly speaking, the logs mainly record the recommendation context, the user's behavior sequence, and the deployed behavior policy at that time. 
The recommendation context and user's behavior sequence are used to form the user and item features. 
The recorded behavior policy is used to reproduce the probability of each action at that time for counterfactual policy evaluation.
As shown in Figure \ref{Fig:RL_Ranking_System}, the recommendation engine contains five major components: an RS agent, an item pool, a logging center, a training component, and an online KV system. The workflow of our system mainly consists of two loops: the online planning loop and the online learning loop. 
As is shown in the left bottom of Figure \ref{Fig:RL_Ranking_System}, the online planning loop is the loop of the interactions between the recommendation agent and users. We will record the item information of each session and the user's behavior for each item. The second is a learning loop (on the right in Figure \ref{Fig:RL_Ranking_System}) where the training process happens. 
Whenever the model is updated, it will be rewritten to the online KV system. We will record the network architecture and network parameters at that time. 
The two working loops are connected through the logging center and the online KV system.
We will record the corresponding user and item features stored in the online KV system for each user log.

After aligning the user log, real-time features, and behavior policy, the format of raw logged data is:
\begin{enumerate}[leftmargin=*]
    \item \textbf{Timestamp}: The timestamp when the event happens.
    
    \item \textbf{Session ID}: A unique number that uniquely identifies a user session.
    
    \item \textbf{Sequence ID}: A unique number representing the location of the state in the session (i.e., the page order).
    
    \item \textbf{Exposed items}: A nine-length space-delimited  list representing the nine items exposed to users (left to right top to bottom).
    
    \item \textbf{User feedback}: A nine-length space-delimited  list representing the user's responses to the nine exposed items (left to right top to bottom). The user feedback (0/1) multiplied by the utility of each item will be used as the reward of this slate.
    
    \item \textbf{User Feature}: The anonymized user features include 42-dim user portrait and 64-dim user click history.
    
    \item \textbf{Item Feature}: The 360-dim item features that describe the context of exposed items on this page, such as the item's id, item's category, item embedding, and historical CTR.
    
    \item \textbf{Behavior Policy ID}: A model file that records the network architecture and network parameters of the behavior policy (supervised learning based strategy) at that time. 
\end{enumerate}

\subsection{Data Preprocessing}
The raw logged data need to be further transformed into consecutive pairs of state/action tuples for RL model training.
There are some differences for RS datasets, including the mapping of state to observation, action embedding, action mask and action probability (used in counterfactual policy evaluation). For the sequential slate scenario, there is also a problem of when users leave. Most existing methods directly predict the probability of users leaving at each step, leading to complex MDP state transitions~\cite{hubo}. In this paper, we develop a kind of session padding method instead of modeling the complex state transitions directly. The user's exit behavior can be seen as a situation where the user will have no action on any items next. So we pad the user session to a certain length with zero padding (zero feedback to a random item slate page) to replace the user's exit behavior. More details are listed in the homepage.

\subsection{MDP Formulation}
In this paper, we apply reinforcement learning to RL4RS datasets, in which RL is employed to learn a policy that at each step, selects an item from the item candidates to be recommended. We formulate this recommendation task as a sequential decision problem (i.e., a discrete-time MDP). For the slate recommendation scenario (Dataset A), there are 9 items on one page (i.e., 9-step decision-making needed). For the sequential slate recommendation scenario (Dataset B), we need to make decisions in at most 36 steps, which are 4 item pages. In one episode of solving a user request, the MDP can be described as follows: 

\textbf{States.} A state $s_{t} \in \mathcal{S} $ at each step needs to reflect not only the static user preference but also the dynamic context information. For the former, we represent it by static user features of user portraits and historical user preference for items. The latter is represented by context features, including the features of selected items $a_{i}$ and dynamic statistics up to step $t$. 

\textbf{Actions.} At each state $s_{t}$, an action of the agent is to select an item $a_{t}$ from all valid items $A$ for this step for recommendation. An item can be represented by its ID or its item embedding. We test both discrete and continuous action settings.

\textbf{Transition.} The environment (i.e., the user) gives feedback after all actions are completed, where only the action $a_{t}$ are added into the user context with the others (e.g., user preference) equaling to their current values in $x_{t}$. Accordingly, the next state $s_{t+1}$ is deterministically attained by concatenating $s_{t}$ with the item $a_{t}$.

\textbf{Rewards.} The reward function of users' responses to an item page is defined as $r_{ \text{page}}=r\left(s_{\text{page}}\right)=r\left(s_{0}, a_{0}, a_{1}, \ldots ,a_{\text{page}-1}\right)=\sum_{i=0}^{\text{page}-1} \gamma^{i} p_{i} r_{i}$, i.e., the user's expected utilities over the page, where $\gamma$ is the discount factor, $p_{i}$ denotes the purchase probability of the i-th item under the condition of the whole page, $r_{i}$ denotes the utility of the i-th item. 
Given the step limit $T$, the goal of RL is to maximize the expected cumulative utilities over all episodes, i.e., the sum of utilities over all the item pages.

\textbf{Policy.} The stochastic policy $\pi$ represents a conditional probability distribution over all possible items given a state. Starting from $s_{0}$, it iteratively recommends an item $a_{t}$ based on the state $s_{t}$ at each step $t$, until reaching the step limit $T$ .

\section{Data Understanding Tools}
\label{sec4}
One big challenge of RL-based RS is problem formulation.
It is easy to accidentally prepare data that does not conform to the MDP definition, and applying RL on ill-formulated problems is costly.
Here, we develop a data understanding tool to
check whether the properties of a new RS dataset conform to the RL framework.
The experiment results also show that traditional recommendation scenarios and datasets are not suitable for modeling as MDP problems.
Without the requirement of establishing environment models, this tool is easy to use by simplifying RL as sequence modeling problems and expected rewards as decoded sequence scores. 
Similar ideas have been developed in Trajectory Transformer~\cite{transformer-rl} and Decision Transformer~\cite{transformer-decision}, in which states, actions, and returns are fed into a GPT~\cite{Radford2018ImprovingLU} architecture, and actions are decoded autoregressively.

The tool fits a Transformer-based sequence-to-sequence model~\cite{transformer} on each offline RS dataset.
It encodes the user context features and decodes K items that users may click or buy, which means that the recommendation system will recommend these K items in turn in the next K steps (considering that most RS datasets do not provide a complete page context, here, only one item is recommended at a time, eliminating the slate effect between items within one page). 
We consider using the decode sequence generated by greedy search to represent greedy recommendation (SL-based strategy), and the sequence generated by beam search to represent the recommendation result generated by the optimal RL policy.
The main idea is that we can compare the sequences score of the greedy strategy (even the sequence composed of only hot items which have high immediate reward) with the score of the optimal sequence to check whether the greedy strategy is good enough.

We build experiments on datasets commonly used in previous studies~\cite{chen2019adversarialgenerative, Bai2019ModelBasedRLnips}, including MovieLens-25m\renewcommand{\thefootnote}{$10$}\footnote{https://grouplens.org/datasets/movielens}, 
RecSys15 YooChoose\renewcommand{\thefootnote}{$11$}\footnote{https://recsys.acm.org/recsys15/challenge/}, 
Last.fm\renewcommand{\thefootnote}{$12$}\footnote{http://ocelma.net/MusicRecommendationDataset/lastfm-1K.html},
CIKMCup2016\renewcommand{\thefootnote}{$13$}\footnote{https://competitions.codalab.org/competitions/11161}
and RL4RS-Slate (RL4RS-SeqSlate has the same result under this experiment setting). 
Without losing generality, we only consider the following trajectories within 5 steps. For each dataset, we choose 10000 users randomly as test set, and others as train set. For each user, we calculate the greedy search result and the top 100 item sequences (beam search width as 100).
The results are shown in Table \ref{table:mdp2}. Using the beam-search width as 100, we report the averaged score of the first 5\% quantile (top 5) decode sequence, the first 20\% quantile (top 20) sequence, and the sequence generated by the greedy strategy. We also calculate the averaged score of the first 5\% quantile and 20\% quantile sequences when limiting the candidate items to hot items (100 items with the highest probability at the first decode step). All these scores are normalized by the average score of the first 5\% sequence for each dataset.
It can be seen that there is a significant gap in the scores between the best 5\% item sequences and the greedy strategy for the RL4RS dataset (the greedy strategy scored 0.62 relatives to the best 5\% sequences), let alone the sequences composed of hot items which scored 0.01 score. 
The result shows that it is necessary to model the RL4RS dataset as an RL problem. 
On the contrary, the score of greedy strategies for traditional RS datasets is very close to that of the best RL policies (the first 5\% decode sequence), which are 1.25 for RecSys15 and 0.99 for MovieLens, 1.09 for Last.fm, and 4.5 for CIKMCup2016.
It indicates that the greedy strategy (i.e., sequential recommendation methods) is well enough and shows the shortcomings of benchmarking on datasets collected from traditional scenarios. 
In practice, the data understanding tool may help us understand the dataset in the early stages, catch improperly defined RL problems, and lead to a better problem definition (e.g., reward definition).

\begin{table}[tbh]
\centering
\caption{The comparison of average scores of decode sequence between different datasets.}
\label{tab:mdp}
\begin{tabular}{lccccc}
\toprule
Score of & 5\% & 20\% & greedy & hot 5\% & hot 20\%\\ \hline
RecSys15 & 1.00 & 0.53 & 1.26  & 0.81 & 0.42 \\
MovieLens & 1.00 & 0.64 & 0.99  & 0.98 & 0.62  \\
Last.fm & 1.00 & 0.47 & 1.09  & 0.90 & 0.49  \\
CIKMCup2016 & 1.00 & 0.30 & 4.50  & 0.99 & 0.28  \\
RL4RS-Slate & 1.00 & 0.76 & 0.62 & 0.01 & 0.01  \\
\bottomrule
\end{tabular}
\label{table:mdp2}
\vspace{-4mm}
\end{table}

\begin{table*}[htb]
\centering
\caption{The experiment results of simulation environment construction. 
}
\label{tab:sl_training}
\begin{tabular}{@{\extracolsep{4pt}}llccccccc}
\toprule   
{} & {} & {Slate-wise Classification} & \multicolumn{2}{c}{Item-wise Classification}  & \multicolumn{4}{c}{Item-wise Rank}\\
 \cmidrule{3-3} 
 \cmidrule{4-5} 
 \cmidrule{6-9} 
 Dataset & Method & Accuracy & AUC & Accuracy & AUC & Precision & Recall & F1 score \\ 
\midrule
RL4RS-Slate  & DNN & 0.386 & 0.884 & 0.799 & 0.876 & 0.826 & 0.843 & 0.834 \\ 
  & Wide\&Deep & 0.398 & 0.897 & 0.812 & 0.894 & 0.827 & 0.875 & 0.850 \\ 
  & GRU4Rec & 0.429 & \textbf{0.913} & \textbf{0.827} & \textbf{0.907} & \textbf{0.862} & 0.846 & 0.854 \\ 
  & DIEN & \textbf{0.430} & 0.911 & 0.826 & \textbf{0.907} & \textbf{0.862} & \textbf{0.848} & \textbf{0.855} \\ 
  & Adversarial & - & - & - & 0.880 & 0.830 & 0.838 & 0.834 \\ 
RL4RS-SeqSlate  & DNN & 0.683 & 0.945 & 0.858 & 0.935 & 0.759 & 0.727 & 0.743 \\ 
  & Wide\&Deep & 0.694 & 0.951 & 0.871 & 0.939 & 0.761 & 0.772 & 0.766 \\ 
  & GRU4Rec & 0.721 & \textbf{0.969} & 0.898 & \textbf{0.961} & \textbf{0.829} & 0.785 & \textbf{0.806} \\ 
  & DIEN & \textbf{0.722} & 0.965 & \textbf{0.890} & 0.960 & 0.820 & \textbf{0.791} & 0.805 \\ 
  & Adversarial & - & - & - & 0.926 & 0.742 & 0.741 & 0.741 \\ 
\bottomrule
\end{tabular}
\end{table*}

\begin{table*}[htb]
\centering
\caption{Reward estimation performance comparison between different supervised learning methods for environment simulation (mean., abs., and std. represent the mean, absolute mean, and standard deviation of reward prediction error respectively).}
\begin{tabular}{lcccccl}\toprule
& \multicolumn{3}{c}{RL4RS-Slate} & \multicolumn{3}{c}{RL4RS-SeqSlate}
\\\cmidrule(lr){2-4}\cmidrule(lr){5-7}
           & mean.  & abs. & std.    & mean.  & abs. & \quad std. \\\midrule
DNN    & 0.9 ($\pm$6.1) & 41.6 ($\pm$2.8) & 66.3 ($\pm$5.1) & 5.4 ($\pm$20.2) & 89.0 ($\pm$9.7) & 118.7 ($\pm$3.3) \\
Wide\&Deep & 1.0 ($\pm$4.3) & 41.1 ($\pm$2.5) & 64.9 ($\pm$4.4) & 6.8 ($\pm$16.5) & 78.7 ($\pm$7.3) & 110.9 ($\pm$3.2) \\
GRU4Rec & -1.2 ($\pm$1.9) & 36.5 ($\pm$2.2) & 61.7 ($\pm$4.3) & 3.3 ($\pm$13.2) & 68.9 ($\pm$6.5) & \textbf{102.5 ($\pm$2.8)}\\
DIEN   & \textbf{0.5 ($\pm$2.3)} & \textbf{36.2 ($\pm$2.2)} & \textbf{60.9 ($\pm$3.8)} & \textbf{-2.4 ($\pm$12.9)} & \textbf{66.2 ($\pm$7.1)} & 102.8 ($\pm$2.8) \\\bottomrule
\end{tabular}
\label{table:simulator_eval}
\vspace{-4mm}
\end{table*}

\section{Evaluation Framework}
\label{sec5}
Unlike academic research, in real applications, it is usually rare to have access to a perfect simulator. A policy should be well evaluated before deployment because policy deployment affects the real world and can be costly.
In applied settings, policy evaluation is not a simple task that can be quantified with only a single indicator (e.g., reward).
As shown in Figure~\ref{Fig:eval}, we attempt to propose a new evaluation framework for RL-based RS, including evaluation of environment simulation, evaluation on simulation environments (online policy evaluation), and counterfactual policy evaluation (offline policy evaluation).
The error bars of all tables mean the standard deviation after running experiments 5 times.
The code for the dataset splits and more experiment results, including evaluation on the test set, are available on GitHub.

\begin{figure}[b]
  \includegraphics[scale=0.16]{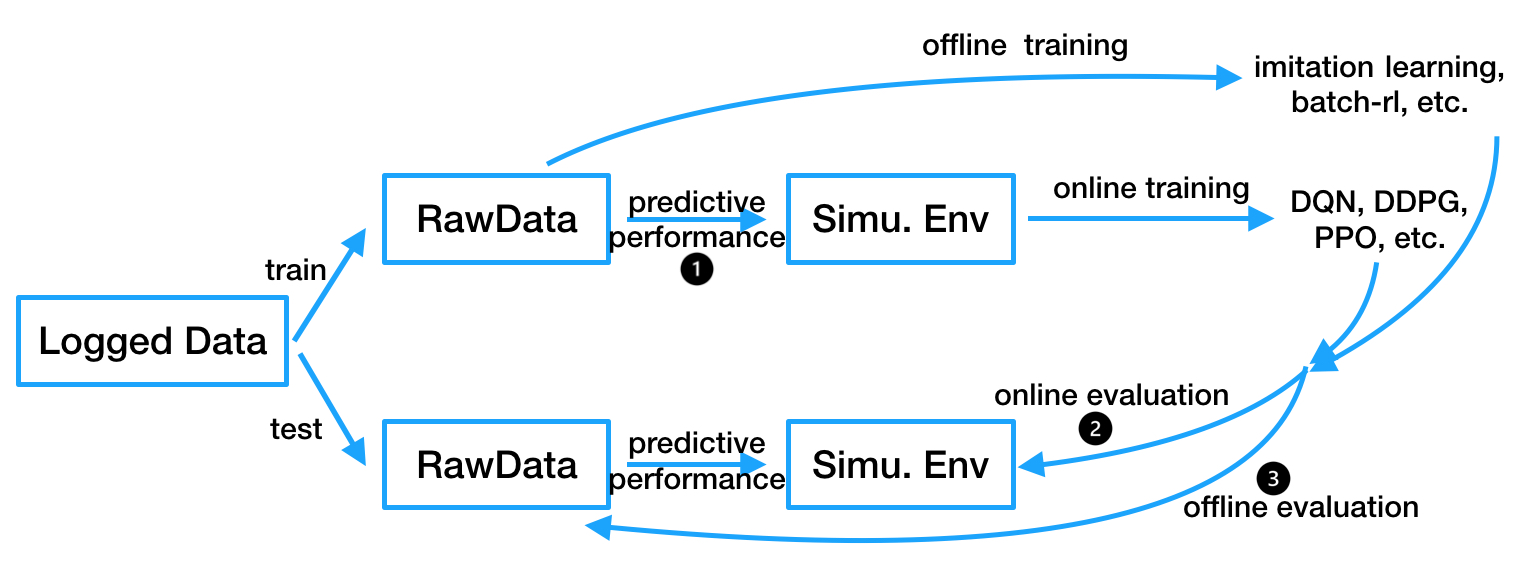}
  \caption{The evaluation framework for RL-based RS tasks. \label{Fig:eval}}
\end{figure}

\subsection{Experiments of Simulation Environment Construction}
A straightforward way to train an RL policy is to interact with pre-trained simulation environments. Most existing RL-based RS works follow this online training manner but introduce a new problem of how to simulate the environment.
For recommender systems, building a simulation environment is to build a perfect user behavior model. Different recommendation scenarios require different user behavior models. For RL4RS datasets, we predict the user's response to item slates and measure the fitting performance of baselines by considering three supervised learning tasks, namely slate-wise classification (multi-class classification), item-wise classification (binary classification), and item-wise rank (ranking task). 

We consider the following supervised learning baselines, including simple DNN, Wide\&Deep~\cite{cheng2016wide}, GRU4Rec~\cite{2015Session}, DIEN~\cite{zhou2019deep} and Adversarial User Model~\cite{chen2019adversarialgenerative}.  
The detailed definition of three supervised learning tasks and network architectures are detailed in the homepage.
From the results shown in Table \ref{tab:sl_training}, GRU4Rec (achieves 0.429 accuracy, 0.913 AUC, and 0.907 AUC in three tasks, respectively) and DIEN (achieves 0.430 accuracy, 0.911 AUC, and 0.907 AUC in three tasks respectively) achieve the best predictive performance. All indicators of the three tasks are represented positive correlations. Note that due to the high purchase rate of this scenario (more than 5 items are purchased per session), the indicators such as item-wise classification accuracy and item-wise rank AUC are much higher than the general RS dataset. 
Because we pad each session to the same length with an all-zero label to simulate the user's exit behavior, the indicators of RL4RS-SeqSlate, such as slate-wise classification accuracy and item-wise classification AUC become higher. Although the prediction of user behavior seems to be accurate enough, we can see from Table \ref{table:simulator_eval} that the estimated reward error is still not low (mean., abs. and std. represent the mean, absolute mean, and standard deviation of reward prediction error respectively). This shows the inconsistency between traditional classification objectives and reward estimation performance, and calls for future research on sophisticated user simulators for RL-based RS.

\subsection{Experiments of Model-free Reinforcement Learning}
Given the learned simulation environment, we can employ any classical RL algorithms. When working on large discrete action spaces (more than thousands of possible actions), one alternative is to address it with a continuous-action RL algorithm and combine policy gradients with a K-NN search~\cite{2015Deep}. 
When only sorting a few candidate items, we can choose to create a variant of DQN called Parametric-Action DQN~\cite{Gauci2018HorizonFO}, PA-DQN in short, in which we input concatenated state-action pairs and output the Q-value for each pair. 
The PA-DQN can make full use of rich item features, which is important for recommendation systems.
Here, the RL4RS suite provides both discrete-action baselines, including Exact-k~\cite{Gong2019ExactKRV}, PG~\cite{Sutton1999PG}, DQN~\cite{dqn}, RAINBOW~\cite{rainbow}, PA-DQN ~\cite{Gauci2018HorizonFO}, A2C~\cite{Mnih2016A2C} and PPO~\cite{schulman2017proximal}, and continuous-action baselines combining policy gradients with a K-NN search, including PG, A2C, DDPG~\cite{ddpg}, TD3~\cite{td3} and PPO. All these algorithms except Exact-k are implemented based on the open-source reinforcement learning library RLlib~\cite{liang2018rllib}. The online evaluation results built on the learned DIEN-based simulation environment are shown in Table \ref{tab:online_training}. 
When modeling the problem as a continuous control RL problem, TD3 achieves the best performance among continuous-action baselines, 
which are  a reward of 178.1 for RL4RS-Slate and a reward of 272.9 for RL4RS-SeqSlate.
In the setting of discrete action space, PG/PPO/A2C (a reward of roughly 160) for RL4RS-Slate and  PPO (a reward of 319.7) for RL4RS-SeqSlate are the best. 
Outside of common RL baselines, we also implement the Exact-K, which is specially designed for the slate recommendation, but its performance is not dominant. Compared with DQN (the rewards of 133.0 for RL4RS-Slate and 226.1 for RL4RS-SeqSlate) and Rainbow (the rewards of 151.5 for RL4RS-Slate and 259.1 for RL4RS-SeqSlate), PA-DQN achieves a similar performance in RL4RS-Slate but a better performance in RL4RS-SeqSlate (a reward of 276.3), which show some advantages of using  concatenated state-action pairs as model input.
Compared with the discrete RL model, the performances of RL baselines are commonly reduced when modeling the problem as a continuous control RL problem, except TD3.
We have prepared detailed reproduction scripts in open-source code. It is convenient for researchers to test more RL baselines in the same experiment setting and make more ablation experiments to figure out why TD3 and PPO are the best for the RL4RS dataset.

\begin{table}[tbh]
\centering
\caption{Performance comparison between model-free RL algorithms on the learned simulation environment.}
\label{tab:online_training}
\begin{tabular}{lcc}
\toprule
Reward of  & RL4RS-Slate & RL4RS-SeqSlate \\ \hline
PG & \textbf{161.1($\pm$2.6)} & 275.8($\pm$2.3)   \\
DQN & 133.0($\pm$2.3) & 226.1($\pm$3.5)   \\
RAINBOW & 151.5($\pm$2.8) & 259.1($\pm$3.3)   \\
PA-DQN & 150.1($\pm$5.1) & 276.3($\pm$6.2)  \\
A2C & 159.8($\pm$4.5) & 282.1($\pm$9.2)   \\
PPO & 160.4($\pm$6.0) & \textbf{319.7($\pm$10.8)}  \\
Exact-k & 136.8($\pm$3.1) & -  \\ \hline
PG-conti & 116.1($\pm$1.2) & 102.8($\pm$3.1)   \\
A2C-conti & 120.3($\pm$4.9) & 218.5($\pm$8.6)   \\
PPO-conti & 127.5($\pm$3.3) & 261.9($\pm$4.8)  \\
DDPG & 146.1($\pm$3.1) & 240.9($\pm$5.4)  \\
TD3 & \textbf{178.1($\pm$3.6)} & \textbf{272.9($\pm$6.2)}  \\
\bottomrule
\end{tabular}
\label{table:env}
\vspace{-4mm}
\end{table}


\subsection{Experiments of Batch Reinforcement Learning}
Offline policy learning aims to learn a better policy from the offline experience pool collected by poor behavior policies. Corresponding to RL-based RS, we have the user feedback data collected by supervised learning based behavior policies as the offline dataset.
RL4RS suite provides three kinds of baselines, imitation learning (behavioral cloning), model-free batch RL (BCQ~\cite{fujimoto2019off} and CQL~\cite{2020Conservative}), and model-based batch RL (MOPO~\cite{mopo} and COMBO~\cite{yu2021combo}).
All algorithms are implemented based on the open-source batch RL library d3rlpy~\cite{seno2020d3rlpy}. 

The online evaluation results built on the learned DIEN-based simulation  environment are shown in Table \ref{tab:offline_training}. 
Because the current open-source MOPO and COMBO algorithms only support continuous actions, we test the two algorithms in the simulation environment under a continuous action space setting.
It can be seen that the BC algorithm (a reward of 223.9 ) which imitates behavior policies achieves higher rewards than the averaged rewards per session (a reward of 164.5) in the RL4RS-SeqSlate dataset. This may be due to BC taking the best item per step rather than sampling according to probability.
In addition, BCQ and CQL achieve much better results than BC, which shows the effectiveness of Batch RL. Compared with model-free RL baselines, batch RL methods achieve lower rewards, which is reasonable because model-free RL is trained on the test environment in an online manner. 
Unlike model-free RL baselines, there is a big performance difference here for batch RL models, especially for CQL (achieves a reward of 103.5 for discrete action space and reward of 82.3 for continuous action space).
It may be because the continuous action space is more difficult to fit for offline batched data, while in simulation environments, the continuous action is first converted into discrete item ID and then used to predict user feedback (i.e., environmental rewards).
As for MOPO-conti and COMBO-conti, they achieve bad performance compared with BCQ-conti and CQL-conti under the same continuous action space setting. In the training log, we find that the dynamic models of MOPO-conti and COMBO-conti are trained poorly. Unlike simulation environments, which only predicts user feedback, the dynamic models directly predict the next observation and reward from the offline dataset, which leads to more serious predictive difficulty. It may be one of the reasons for the poor performance of model-based batch RL.
More in-depth analysis and ablation experiments as left as future work.

\begin{table}[tb]
\centering
\caption{Performance comparison between batch RL algorithms on the same learned simulation environment.}
\label{tab:offline_training}
\begin{tabular}{lcc}
\toprule
Reward of & RL4RS-Slate & RL4RS-SeqSlate \\ \hline
BC & 94.1($\pm$1.6) & 223.9($\pm$4.5) \\
BCQ & \textbf{127.4}($\pm$3.6) & \textbf{262.4}($\pm$5.7) \\
CQL & 103.5($\pm$3.1) & 247.5($\pm$5.3) \\
BCQ-conti & 109.1($\pm$4.6) & 229.0($\pm$6.2) \\
CQL-conti & 82.3($\pm$3.9) & 199.5($\pm$6.5) \\
MOPO-conti & 42.7($\pm$4.3) & 100.3($\pm$6.1) \\
COMBO-conti & 68.9($\pm$4.2) & 127.5($\pm$5.9) \\
\bottomrule
\end{tabular}
\label{table:offline_training}
\end{table}

\subsection{Counterfactual Policy Evaluation}
Counterfactual policy evaluation (CPE) is a set of methods used to predict the performance of a newly learned policy without having to deploy it online~\cite{is,dudik2011doubly,jiang2016doubly,thomas2016magic}. 
The RL4RS suite provides several well-known counterfactual policy evaluation (CPE) methods to score trained models offline, including importance sampling~\cite{is} (IS),  step-wise weighted importance sampling (SWIS), doubly-robust~\cite{dudik2011doubly} (DR), and sequential doubly-robust~\cite{jiang2016doubly} (Sequential DR).

We use four counterfactual policy evaluation methods (IS, SWIS, DR, and sequential DR) to evaluate the trained RL agent described in Sections 5.2 and 5.3. The results are shown in Table \ref{tab:offline_evaluation}.
The CPE results indicate the expected performance of the newly trained policy relative to the policy that generated the dataset. 
A score of 1.0 means that the RL and the logged policy match in performance. For example, the DR score of DQN (1.43) implies that the DQN policy should achieve roughly 1.43x as much cumulative reward as the behavior policy.  The two IS-based methods are relatively unstable in value. After multiplying the propensity ratio (i.e., action probability of trained policy divided by the probability of logged policy) of each step, it is easy to reach an unreasonable value. When the SWIS method clips the propensity ratio of each step to $[0.1, 10]$, almost all samples reach the clipped limit. The DR-based methods perform relatively well, thanks to their stable state value estimation equipped with the learned simulation environment. 
Compared with model-free RL (PPO with scores of 1.24 and 1.45), Batch RL (CQL with scores of 2.20 and 2.95) achieves higher evaluation scores under the DR-based methods.
Although the results of CPE methods seem not convincing,
at least they draw a different conclusion from the evaluation on the simulation environment.
We will explore more state-of-the-art CPE methods in the future.

\begin{table}[h]
\small
\centering
\caption{Counterfactual policy evaluation on learned model-free RL and batch RL algorithms.}
\label{tab:offline_evaluation}
\begin{tabular}{lcccc}
\toprule
RL4RS-Slate  & IS & SWIS & DR & Seq. DR \\ \hline
BCQ & 9.12($\pm$0.61) & 1.00($\pm$0.00) & 1.55($\pm$0.03) & 1.59($\pm$0.06) \\
CQL & \textbf{130.61($\pm$6.32)} & \textbf{1.41($\pm$0.94)} & \textbf{2.20($\pm$0.02)} & \textbf{2.95($\pm$0.04)} \\
DQN & 9.26($\pm$0.97) & 1.00($\pm$0.00) & 1.43($\pm$0.02) & 1.66($\pm$0.08) \\
PPO & 9.150($\pm$1.23) & 1.00($\pm$0.00) & 1.24($\pm$0.03) & 1.45($\pm$0.07) \\
A2C & 9.62($\pm$1.12) & 1.00($\pm$0.00) & 0.77($\pm$0.02) & 0.93($\pm$0.03) \\
\bottomrule
\end{tabular}
\label{table:offline_training2}
\end{table}

\section{Live Experiments}
Although we have tested RL methods using two offline evaluation methods (evaluating on simulation environments and counterfactual policy evaluation), the most accurate way is to deploy them online.
We apply DIEN, A2C, DQN, and PPO to the online recommendation application described in Section 3.
DIEN works as the greedy recommendation strategy, which is also the main deployed algorithm of this scenario previously.
The RL methods are trained on the learned simulation environment (described in Section 5.1) under the sequential slate recommendation setting (described in Section 3), and are updated weekly.
We employ metrics including IPV (Individual Page View), CVR (Conversion Rate), and GMV (Gross Merchandise Volume).

During the A/B test in one month, the averaged results are presented in Table 9. 
On each day of the test, the PPO algorithm can lead to more GMV than the DIEN, A2C, and DQN (an increase of $23.9\%, 8.4\%, and 12.3\%$, respectively).
Compared with the greedy recommendation strategy, which recommends only the items that users are interested in at each step, RL methods can make full use of the multi-step decision-making characteristics of this scenario, and minimize the loss of CVR while mixing the recommended items users are interested in and other items.
In addition, the performance ranking of online deployment (PPO>A2C>DQN>SL) is consistent with the ranking in the simulation environment (PPO>A2C>DQN>SL). It shows that offline experiments can replace live experiments to a certain extent.
Here, we promise that we will help researchers who achieve better offline performance deploy their algorithms online for accurate evaluation, as we did earlier in IEEE BigData Cup 2021\renewcommand{\thefootnote}{$14$}\footnote{http://bigdataieee.org/BigData2021/BigDataCupChallenges.html}.

\begin{table}[tbh]
\caption{Comparison between supervised learning strategy and RL baselines in the live experiments.}
\label{tab:ec}
\centering
{
\begin{tabular}{l c c c}
\toprule
\textbf{Method} & IPV & CVR & GMV\\
\midrule
DIEN & 1.000 & \textbf{1.000}  & 1.000 \\
A2C & \textbf{1.206} & 0.801 & 1.190 \\
DQN & 0.966 & 0.874  & 1.143 \\
PPO  & 1.133 & 0.853 &  \textbf{1.239}\\
\bottomrule\\
\end{tabular}}
\end{table}

\section{Availability}
RL4RS datasets are available for download from Zenodo \textbf{ \textit {\url{https://zenodo.org/record/6622390##.YqBBpRNBxQK}}} or Github \textbf{ \textit {\url{https://github.com/fuxiAIlab/RL4RS}}}.
The dataset DOI is \textbf{ \textit{10.5281/zenodo.6622390}}.
The meta-data page using Web standards is \textbf{ \textit {\url{https://github.com/fuxiAIlab/RL4RS/blob/main/index.html}}}.
The corresponding code for dataset processing and model training is maintained at \textbf{ \textit {\url{https://github.com/fuxiAIlab/RL4RS}}}.
To support findability and sustainability, the RL4RS suit is published as an online resource at \textbf{ \textit {\url{https://fuxi-up-research.gitbook.io/fuxi-up-challenges/dataset/rl4rs}}}. 
Documentation includes all relevant metadata specified to the research community and users. It is freely accessible under the \textit{Creative Commons Attribution 4.0 International license}, making it reusable for almost any purpose. The authors bear all responsibility in case of violation of rights, and confirm the CC licenses for the included datasets.

\subsection{Updating and Reusability}
RL4RS is supported by a team of researchers from the Fuxi AI Lab, the Netease Inc.  
RL4RS will be supported and maintained for three years.
An older and smaller dataset collected from the same application is already in use as a contribution to the data challenge of BigData conference 2021 at \textbf{ \textit {\url{https://www.kaggle.com/c/bigdata2021-rl-recsys/overview}}}. 
In addition to the steps above that make the resource available to the broader community, usage of RL4RS will be promoted to the network of researchers in this project. 
A new dataset for bundle recommendation with variable discounts, flexible recommendation triggers, and modifiable item content is in preparation.
In the future, we will add more datasets from other novel scenarios to support the research of RL-based RS.

\subsection{System Configuration}
Pipelines for all tasks are trained for at most 100 epochs with an early stop on a single GeForce GTX 2080 Ti GPU, 8 CPU, and 64GB memory. 
The code for the dataset splits is also available. 
For all tasks, consistent hyperparameter settings are maintained to enable comparative analysis of the results. The batch size is set at 256 with a learning rate of 0.001 and a probability of 0.1 in dropout layers. The discounted factor is 0.95 and the embedding size is 128.

\section{RL4RS for New Research}
\label{sec6}
The accessibility of raw logged data and separated data before and after RL deployment of RL4RS opens up new exciting research opportunities. Below, we present three promising directions.

\begin{enumerate}[leftmargin=*]
    \item \textbf{Better Environment Simulation Method.}
    Since RL4RS provides raw logged data, it is now possible for researchers to explore the impact of different simulation environment construction methods on RL policy learning. For example, Table 4 shows that the ACC and AUC indicators of the environment model are satisfactory, but the performance of the reward prediction task is not so good. This result is surprising, and provides a strong impetus for additional research into simulation environment construction, such as (a) whether the ACC and AUC metrics that RS commonly used are consistent with the reward prediction error; (b) whether the variance predicted by the environment model should be considered in the learning of RL agent (different user contexts may have different prediction variances).

    \item \textbf{Learning from Offline Data Directly.}  In addition to training RL agents on the environmental model, we can also learn policy directly from offline data to avoid the difficulty of simulation environment construction. Unfortunately, data collected from real-world applications are often generated by myopic behavior policies (e.g., supervised learning based strategies), resulting in ``extrapolation errors'' between the behavior policy and the learned RL policy. Batch RL and Counterfactual Policy Evaluation in the RL-based RS domain have yet to receive much attention, largely because no datasets existed to support the research. By releasing the separated data before and after RL deployment in RL4RS, we hope to foster new interest in this critical area.
    
    \item \textbf{Reconsidering Evaluation Methodology of Applied RL.}  
    Our discovery in the experiments of evaluation on the simulation environment built from test set indicates that the averaged rewards vary a lot under different data split settings, suggesting that the current RL evaluation methodology, i.e., training and evaluating in the same environment, is not yet well-suited for RL-based RS tasks. A comprehensive evaluation methodology should also include evaluation of environment simulation, counterfactual policy evaluation, and evaluation on simulation environments built from test set. A popular evaluation method adopted by most works is to divide users before policy training. Under this setting, although the training and test data are not overlapped, they share the same environment. It means there is still a risk of over-fitting because the estimated reward of the test set will be affected by the train set through the shared environmental model. We believe our results will inspire researchers to reconsider the evaluation methodology of RL-based RS and applied RL.

\end{enumerate}

\section{Related Work}
\label{sec7}
This resource is oriented to the reinforcement learning based recommender systems domain. In 2015, Shani et al.~\cite{shani2005mdp} proposes to model the recommendation system problem as an MDP process for the first time. In 2018, DRN~\cite{zheng2018drn} first apply deep RL to recommendation problems and inspire a series of subsequent works. SlateQ~\cite{Ie2019SlateQAT} is the first work in this domain to concern the slate recommendation scenario. 
Top-k off-policy~\cite{chen2019topk} learns the policy from the offline dataset directly for the first time.
Adversarial User Model~\cite{chen2019adversarialgenerative} first popularizes semi-simulated RS datasets and simulates environments using adversarial learning.
Virtual-Taobao~\cite{Shi2019VirtualTaobaoVR} is the only work to open-source a pre-trained simulation environment (but without raw logged data) that comes from a real application. 
Accordion~\cite{mcinerney2021accordion} implements a trainable Poisson process simulator for interactive systems.
A detailed comparison is listed in Section ~\ref{sec2}.

Outside of the RS domain, research on reinforcement learning for the online environment has an extensive history and we focus here on resources and benchmark tasks where samples are collected from real applications and present in batched offline data, such as DeepMind Control Suite~\cite{2018Control}, D4RL~\cite{2020D4RL} and RL unplugged~\cite{2020RLUnplugged}. All of these resources exclude recommendation scenarios. Most batched datasets of these resources are collected by mixed, nearly random policies, and are usually equipped with a deterministic environment. However, these characteristics are inconsistent with the attribute of real-world data of recommender systems.

\section{Discussion of Limitations}
In this paper, we follow the experiment settings of most related research, i.e.,  training and evaluating the RL agent on the supervised learning based simulator (i.e., the RL environment) built from the user feedback data. However, there are concerns that the pre-training environmental model may not be able to benchmark the model fairly because of the generalization error of the environmental model. We note that such concerns can be partially solved by batch RL algorithms which does not depend on the environmental model, and evaluating the policy on the raw data directly. Due to the limitation of computing resources and reproduction difficulty, we are unable to conduct a comprehensive evaluation of all existing reinforcement learning models under all evaluation settings. However, with the release of source code and leaderboard website, we are expecting researchers to actively submit their models. We are also interested in RS-specific RL exploration, and regard it as an attractive future work.

\section{Conclusion and Future Work}
\label{sec8}
This paper presents the RL4RS resource, the first open-source real-world dataset for RL-based RS. 
For real-world considerations, RL4RS focuses on simulation environment construction, extrapolation error, offline policy learning, and offline policy evaluation, which are ubiquitous and crucial in recommendation scenarios. 
So far, RL4RS has included two novel scenarios: slate recommendation and sequential slate recommendation. The separated data before and after RL deployment are also available for each dataset.
We benchmark some state-of-the-art online and batch RL algorithms under both discrete and continuous action settings. 
In addition, we develop data understanding tools for assembling new datasets easily and propose a new systematic evaluation framework for RL-based RS, including evaluation of environment simulation, counterfactual policy evaluation, and evaluation on simulation environments built from test set (online policy evaluation). 
In the future, we will constantly provide new real-world RS datasets and step further toward real-world RS challenges. We also hope the RL4RS resource will shed some light on future research on applied reinforcement learning and help researchers deploy their own online RL-based recommendation applications.

\newpage
\bibliographystyle{abbrvnat}
\bibliography{main}

\begin{thebibliography}{45}
\providecommand{\natexlab}[1]{#1}
\providecommand{\url}[1]{\texttt{#1}}
\expandafter\ifx\csname urlstyle\endcsname\relax
  \providecommand{\doi}[1]{doi: #1}\else
  \providecommand{\doi}{doi: \begingroup \urlstyle{rm}\Url}\fi

\bibitem[Bai et~al.(2019)Bai, Guan, and Wang]{Bai2019ModelBasedRLnips}
X.~Bai, J.~Guan, and H.~Wang.
\newblock Model-based reinforcement learning with adversarial training for
  online recommendation.
\newblock In \emph{NeurIPS}, 2019.

\bibitem[Chen et~al.(2021)Chen, Kevin, Aravind, Kimin, Aditya, Michael, Pieter,
  Aravind, and Igor]{transformer-decision}
L.~Chen, L.~Kevin, R.~Aravind, L.~Kimin, G.~Aditya, L.~Michael, A.~Pieter,
  S.~Aravind, and M.~Igor.
\newblock Decision transformer: Reinforcement learning via sequence modeling.
\newblock \emph{arXiv preprint arXiv:2106.01345}, 2021.

\bibitem[Chen et~al.(2019{\natexlab{a}})Chen, Beutel, Covington, Jain,
  Belletti, and Chi]{chen2019topk}
M.~Chen, A.~Beutel, P.~Covington, S.~Jain, F.~Belletti, and E.~H. Chi.
\newblock Top-k off-policy correction for a reinforce recommender system.
\newblock In \emph{Proceedings of the Twelfth ACM International Conference on
  Web Search and Data Mining}, pages 456--464, 2019{\natexlab{a}}.

\bibitem[Chen et~al.(2019{\natexlab{b}})Chen, Li, Li, Jiang, Qi, and
  Song]{chen2019adversarialgenerative}
X.~Chen, S.~Li, H.~Li, S.~Jiang, Y.~Qi, and L.~Song.
\newblock Generative adversarial user model for reinforcement learning based
  recommendation system.
\newblock In \emph{International Conference on Machine Learning}, pages
  1052--1061. PMLR, 2019{\natexlab{b}}.

\bibitem[Cheng et~al.(2016)Cheng, Koc, Harmsen, Shaked, Chandra, Aradhye,
  Anderson, Corrado, Chai, Ispir, et~al.]{cheng2016wide}
H.-T. Cheng, L.~Koc, J.~Harmsen, T.~Shaked, T.~Chandra, H.~Aradhye,
  G.~Anderson, G.~Corrado, W.~Chai, M.~Ispir, et~al.
\newblock Wide \& deep learning for recommender systems.
\newblock In \emph{Proceedings of the 1st workshop on deep learning for
  recommender systems}, pages 7--10, 2016.

\bibitem[Dud{\'\i}k et~al.(2011)Dud{\'\i}k, Langford, and Li]{dudik2011doubly}
M.~Dud{\'\i}k, J.~Langford, and L.~Li.
\newblock Doubly robust policy evaluation and learning.
\newblock \emph{arXiv preprint arXiv:1103.4601}, 2011.

\bibitem[Dulac-Arnold et~al.(2015)Dulac-Arnold, Evans, van Hasselt, Sunehag,
  Lillicrap, Hunt, Mann, Weber, Degris, and Coppin]{2015Deep}
G.~Dulac-Arnold, R.~Evans, H.~van Hasselt, P.~Sunehag, T.~Lillicrap, J.~Hunt,
  T.~Mann, T.~Weber, T.~Degris, and B.~Coppin.
\newblock Deep reinforcement learning in large discrete action spaces.
\newblock \emph{arXiv preprint arXiv:1512.07679}, 2015.

\bibitem[Fu et~al.(2020)Fu, Kumar, Nachum, Tucker, and Levine]{2020D4RL}
J.~Fu, A.~Kumar, O.~Nachum, G.~Tucker, and S.~Levine.
\newblock D4rl: Datasets for deep data-driven reinforcement learning.
\newblock \emph{arXiv}, 2020.

\bibitem[Fujimoto et~al.(2018)Fujimoto, Hoof, and Meger]{td3}
S.~Fujimoto, H.~Hoof, and D.~Meger.
\newblock Addressing function approximation error in actor-critic methods.
\newblock In \emph{International conference on machine learning}, pages
  1587--1596. PMLR, 2018.

\bibitem[Fujimoto et~al.(2019)Fujimoto, Meger, and Precup]{fujimoto2019off}
S.~Fujimoto, D.~Meger, and D.~Precup.
\newblock Off-policy deep reinforcement learning without exploration.
\newblock In \emph{International Conference on Machine Learning}, pages
  2052--2062, 2019.

\bibitem[Gauci et~al.(2018)Gauci, Conti, Liang, Virochsiri, He, Kaden,
  Narayanan, Ye, and Fujimoto]{Gauci2018HorizonFO}
J.~Gauci, E.~Conti, Y.~Liang, K.~Virochsiri, Y.~R. He, Z.~Kaden, V.~Narayanan,
  X.~Ye, and S.~Fujimoto.
\newblock Horizon: Facebook's open source applied reinforcement learning
  platform.
\newblock \emph{ArXiv}, abs/1811.00260, 2018.

\bibitem[Gebru et~al.(2018)Gebru, Morgenstern, Vecchione, Vaughan, Wallach,
  Daum{\'e}~III, and Crawford]{datasheet}
T.~Gebru, J.~Morgenstern, B.~Vecchione, J.~W. Vaughan, H.~Wallach,
  H.~Daum{\'e}~III, and K.~Crawford.
\newblock Datasheets for datasets.
\newblock \emph{arXiv preprint arXiv:1803.09010}, 2018.

\bibitem[Gong et~al.(2019)Gong, Zhu, Duan, Liu, Guan, Sun, Ou, and
  Zhu]{Gong2019ExactKRV}
Y.~Gong, Y.~Zhu, L.~Duan, Q.~Liu, Z.~Guan, F.~Sun, W.~Ou, and K.~Q. Zhu.
\newblock Exact-k recommendation via maximal clique optimization.
\newblock \emph{Proceedings of the 25th ACM SIGKDD International Conference on
  Knowledge Discovery \& Data Mining}, 2019.

\bibitem[Gulcehre et~al.(2020)Gulcehre, Wang, Novikov, Paine, and
  Freitas]{2020RLUnplugged}
C.~Gulcehre, Z.~Wang, A.~Novikov, T.~L. Paine, and N.~D. Freitas.
\newblock Rl unplugged: Benchmarks for offline reinforcement learning.
\newblock \emph{arXiv}, 2020.

\bibitem[Hessel et~al.(2018)Hessel, Modayil, Van~Hasselt, Schaul, Ostrovski,
  Dabney, Horgan, Piot, Azar, and Silver]{rainbow}
M.~Hessel, J.~Modayil, H.~Van~Hasselt, T.~Schaul, G.~Ostrovski, W.~Dabney,
  D.~Horgan, B.~Piot, M.~Azar, and D.~Silver.
\newblock Rainbow: Combining improvements in deep reinforcement learning.
\newblock In \emph{Thirty-second AAAI conference on artificial intelligence},
  2018.

\bibitem[Hidasi et~al.(2015)Hidasi, Karatzoglou, Baltrunas, and
  Tikk]{2015Session}
B.~Hidasi, A.~Karatzoglou, L.~Baltrunas, and D.~Tikk.
\newblock Session-based recommendations with recurrent neural networks.
\newblock \emph{Computer ence}, 2015.

\bibitem[Horvitz and Thompson(1952)]{is}
D.~G. Horvitz and D.~J. Thompson.
\newblock A generalization of sampling without replacement from a finite
  universe.
\newblock \emph{Journal of the American statistical Association}, 47\penalty0
  (260):\penalty0 663--685, 1952.

\bibitem[Hu et~al.(2018)Hu, Da, Zeng, Yu, and Xu]{hubo}
Y.~Hu, Q.~Da, A.~Zeng, Y.~Yu, and Y.~Xu.
\newblock Reinforcement learning to rank in e-commerce search engine:
  Formalization, analysis, and application.
\newblock In \emph{Proceedings of the 24th ACM SIGKDD International Conference
  on Knowledge Discovery \& Data Mining}, pages 368--377, 2018.

\bibitem[Huzhang et~al.(2021)Huzhang, Pang, Gao, Liu, Shen, Zhou, Da, Zeng, Yu,
  Yu, et~al.]{2020Validation}
G.~Huzhang, Z.~Pang, Y.~Gao, Y.~Liu, W.~Shen, W.-J. Zhou, Q.~Da, A.~Zeng,
  H.~Yu, Y.~Yu, et~al.
\newblock Aliexpress learning-to-rank: Maximizing online model performance
  without going online.
\newblock \emph{IEEE Transactions on Knowledge and Data Engineering}, 2021.

\bibitem[Ie et~al.(2019{\natexlab{a}})Ie, Hsu, Mladenov, Jain, Narvekar, Wang,
  Wu, and Boutilier]{ie2019recsim}
E.~Ie, C.-w. Hsu, M.~Mladenov, V.~Jain, S.~Narvekar, J.~Wang, R.~Wu, and
  C.~Boutilier.
\newblock Recsim: A configurable simulation platform for recommender systems.
\newblock \emph{arXiv preprint arXiv:1909.04847}, 2019{\natexlab{a}}.

\bibitem[Ie et~al.(2019{\natexlab{b}})Ie, Jain, Wang, Narvekar, Agarwal, Wu,
  Cheng, Chandra, and Boutilier]{Ie2019SlateQAT}
E.~Ie, V.~Jain, J.~Wang, S.~Narvekar, R.~Agarwal, R.~Wu, H.-T. Cheng,
  T.~Chandra, and C.~Boutilier.
\newblock Slateq: A tractable decomposition for reinforcement learning with
  recommendation sets.
\newblock In \emph{IJCAI}, 2019{\natexlab{b}}.

\bibitem[Janner et~al.(2021)Janner, Li, and Levine]{transformer-rl}
M.~Janner, Q.~Li, and S.~Levine.
\newblock Reinforcement learning as one big sequence modeling problem.
\newblock \emph{arXiv preprint arXiv:2106.02039}, 2021.

\bibitem[Jiang and Li(2016)]{jiang2016doubly}
N.~Jiang and L.~Li.
\newblock Doubly robust off-policy value evaluation for reinforcement learning.
\newblock In \emph{International Conference on Machine Learning}, pages
  652--661. PMLR, 2016.

\bibitem[Kumar et~al.(2020)Kumar, Zhou, Tucker, and Levine]{2020Conservative}
A.~Kumar, A.~Zhou, G.~Tucker, and S.~Levine.
\newblock Conservative q-learning for offline reinforcement learning.
\newblock \emph{ArXiv}, 2020.

\bibitem[Liang et~al.(2018)Liang, Liaw, Nishihara, Moritz, Fox, Goldberg,
  Gonzalez, Jordan, and Stoica]{liang2018rllib}
E.~Liang, R.~Liaw, R.~Nishihara, P.~Moritz, R.~Fox, K.~Goldberg, J.~E.
  Gonzalez, M.~I. Jordan, and I.~Stoica.
\newblock {RLlib}: Abstractions for distributed reinforcement learning.
\newblock In \emph{International Conference on Machine Learning ({ICML})},
  2018.

\bibitem[Lillicrap et~al.(2015)Lillicrap, Hunt, Pritzel, Heess, Erez, Tassa,
  Silver, and Wierstra]{ddpg}
T.~P. Lillicrap, J.~J. Hunt, A.~Pritzel, N.~Heess, T.~Erez, Y.~Tassa,
  D.~Silver, and D.~Wierstra.
\newblock Continuous control with deep reinforcement learning.
\newblock \emph{arXiv preprint arXiv:1509.02971}, 2015.

\bibitem[McInerney et~al.(2021)McInerney, Elahi, Basilico, Raimond, and
  Jebara]{mcinerney2021accordion}
J.~McInerney, E.~Elahi, J.~Basilico, Y.~Raimond, and T.~Jebara.
\newblock Accordion: a trainable simulator for long-term interactive systems.
\newblock In \emph{Proceedings of the 15th ACM Conference on Recommender
  Systems}, pages 102--113, 2021.

\bibitem[Mnih et~al.(2013)Mnih, Kavukcuoglu, Silver, Graves, Antonoglou,
  Wierstra, and Riedmiller]{dqn}
V.~Mnih, K.~Kavukcuoglu, D.~Silver, A.~Graves, I.~Antonoglou, D.~Wierstra, and
  M.~Riedmiller.
\newblock Playing atari with deep reinforcement learning.
\newblock \emph{arXiv preprint arXiv:1312.5602}, 2013.

\bibitem[Mnih et~al.(2016)Mnih, Badia, Mirza, Graves, Lillicrap, Harley,
  Silver, and Kavukcuoglu]{Mnih2016A2C}
V.~Mnih, A.~P. Badia, M.~Mirza, A.~Graves, T.~P. Lillicrap, T.~Harley,
  D.~Silver, and K.~Kavukcuoglu.
\newblock Asynchronous methods for deep reinforcement learning.
\newblock In \emph{ICML}, 2016.

\bibitem[Radford and Narasimhan(2018)]{Radford2018ImprovingLU}
A.~Radford and K.~Narasimhan.
\newblock Improving language understanding by generative pre-training.
\newblock 2018.

\bibitem[Rivest and Dusse(1992)]{rivest1992md5}
R.~Rivest and S.~Dusse.
\newblock The md5 message-digest algorithm, 1992.

\bibitem[Rohde et~al.(2018)Rohde, Bonner, Dunlop, Vasile, and
  Karatzoglou]{Rohde2018RecoGymAR}
D.~Rohde, S.~Bonner, T.~Dunlop, F.~Vasile, and A.~Karatzoglou.
\newblock Recogym: A reinforcement learning environment for the problem of
  product recommendation in online advertising.
\newblock \emph{ArXiv}, abs/1808.00720, 2018.

\bibitem[Schulman et~al.(2017)Schulman, Wolski, Dhariwal, Radford, and
  Klimov]{schulman2017proximal}
J.~Schulman, F.~Wolski, P.~Dhariwal, A.~Radford, and O.~Klimov.
\newblock Proximal policy optimization algorithms.
\newblock \emph{arXiv preprint arXiv:1707.06347}, 2017.

\bibitem[Seno(2020)]{seno2020d3rlpy}
T.~Seno.
\newblock d3rlpy: An offline deep reinforcement library.
\newblock \url{https://github.com/takuseno/d3rlpy}, 2020.

\bibitem[Shani et~al.(2005)Shani, Heckerman, Brafman, and
  Boutilier]{shani2005mdp}
G.~Shani, D.~Heckerman, R.~I. Brafman, and C.~Boutilier.
\newblock An mdp-based recommender system.
\newblock \emph{Journal of Machine Learning Research}, 6\penalty0 (9), 2005.

\bibitem[Shi et~al.(2019)Shi, Yu, Da, Chen, and Zeng]{Shi2019VirtualTaobaoVR}
J.~Shi, Y.~Yu, Q.~Da, S.-Y. Chen, and A.~Zeng.
\newblock Virtual-taobao: Virtualizing real-world online retail environment for
  reinforcement learning.
\newblock In \emph{AAAI}, 2019.

\bibitem[Sutton et~al.(1999)Sutton, McAllester, Singh, and
  Mansour]{Sutton1999PG}
R.~S. Sutton, D.~A. McAllester, S.~Singh, and Y.~Mansour.
\newblock Policy gradient methods for reinforcement learning with function
  approximation.
\newblock In \emph{NIPS}, 1999.

\bibitem[Tassa et~al.(2018)Tassa, Doron, Muldal, Erez, Li, Casas, Budden,
  Abdolmaleki, Merel, and Lefrancq]{2018Control}
Y.~Tassa, Y.~Doron, A.~Muldal, T.~Erez, Y.~Li, D.~Casas, D.~Budden,
  A.~Abdolmaleki, J.~Merel, and A.~Lefrancq.
\newblock Deepmind control suite.
\newblock \emph{arXiv}, 2018.

\bibitem[Thomas and Brunskill(2016)]{thomas2016magic}
P.~Thomas and E.~Brunskill.
\newblock Data-efficient off-policy policy evaluation for reinforcement
  learning.
\newblock In \emph{International Conference on Machine Learning}, pages
  2139--2148. PMLR, 2016.

\bibitem[Vaswani et~al.(2017)Vaswani, Shazeer, Parmar, Uszkoreit, Jones, Gomez,
  Kaiser, and Polosukhin]{transformer}
A.~Vaswani, N.~Shazeer, N.~Parmar, J.~Uszkoreit, L.~Jones, A.~N. Gomez,
  L.~Kaiser, and I.~Polosukhin.
\newblock Attention is all you need.
\newblock In \emph{Proceedings of the 31st International Conference on Neural
  Information Processing Systems}, NIPS'17, page 6000–6010, Red Hook, NY,
  USA, 2017. Curran Associates Inc.
\newblock ISBN 9781510860964.

\bibitem[Yu et~al.(2020)Yu, Thomas, Yu, Ermon, Zou, Levine, Finn, and Ma]{mopo}
T.~Yu, G.~Thomas, L.~Yu, S.~Ermon, J.~Y. Zou, S.~Levine, C.~Finn, and T.~Ma.
\newblock Mopo: Model-based offline policy optimization.
\newblock \emph{Advances in Neural Information Processing Systems},
  33:\penalty0 14129--14142, 2020.

\bibitem[Yu et~al.(2021)Yu, Kumar, Rafailov, Rajeswaran, Levine, and
  Finn]{yu2021combo}
T.~Yu, A.~Kumar, R.~Rafailov, A.~Rajeswaran, S.~Levine, and C.~Finn.
\newblock Combo: Conservative offline model-based policy optimization.
\newblock \emph{Advances in neural information processing systems},
  34:\penalty0 28954--28967, 2021.

\bibitem[Zhao et~al.(2017)Zhao, Zhang, Xia, Ding, Yin, and Tang]{zhao2017list}
X.~Zhao, L.~Zhang, L.~Xia, Z.~Ding, D.~Yin, and J.~Tang.
\newblock Deep reinforcement learning for list-wise recommendations.
\newblock \emph{arXiv preprint arXiv:1801.00209}, 2017.

\bibitem[Zheng et~al.(2018)Zheng, Zhang, Zheng, Xiang, Yuan, Xie, and
  Li]{zheng2018drn}
G.~Zheng, F.~Zhang, Z.~Zheng, Y.~Xiang, N.~J. Yuan, X.~Xie, and Z.~Li.
\newblock Drn: A deep reinforcement learning framework for news recommendation.
\newblock In \emph{Proceedings of the 2018 World Wide Web Conference}, pages
  167--176, 2018.

\bibitem[Zhou et~al.(2019)Zhou, Mou, Fan, Pi, Bian, Zhou, Zhu, and
  Gai]{zhou2019deep}
G.~Zhou, N.~Mou, Y.~Fan, Q.~Pi, W.~Bian, C.~Zhou, X.~Zhu, and K.~Gai.
\newblock Deep interest evolution network for click-through rate prediction.
\newblock In \emph{Proceedings of the AAAI Conference on Artificial
  Intelligence}, volume~33, pages 5941--5948, 2019.

\end{thebibliography}

\newpage

\appendix

\section{Data Sheet}
\label{appendix_b}
\label{appendix:datasheet}
We follow the documentation frameworks provided by \citet{datasheet}.

\subsection{Motivation}

\paragraph{For what purpose was the dataset created?}

While recently many works claim that they successfully apply reinforcement learning algorithms to recommendation tasks, the experimental settings in these works ($i$) lacks datasets collected from real-world applications, and only test on artificial datasets and semi-simulated datasets with unreasonable  data transformations; ($ii$) often tests in the environment where the RL agent trains on, and lacks uniform and unbiased evaluation criteria. This motivates us to build the first real-world dataset, provide all details from raw data to environment construction, propose a unified and principled evaluation framework, and evaluate to which extent the state-of-the-art models have progressed so far in terms of RL-based RS.


\paragraph{Who created the dataset (e.g., which team, research group) and on behalf of which entity (e.g., company, institution, organization)?} Fuxi AI Lab, the Netease Inc.

\subsection{Composition:}
\paragraph{What do the instances that comprise the dataset represent (e.g., documents, photos, people, countries)? Are there multiple types of instances (e.g., movies, users, and ratings; people and interactions between them; nodes and edges)?} RL4RS contains item metadata and user feedbacks in the dataset.

\paragraph{How many instances are there in total (of each type, if appropriate)?} There are 1719316 user sessions for RL4RS-Slate dataset and 958566 user sessions for RL4RS-SeqSlate dataset.

\paragraph{Does the dataset contain all possible instances or is it a sample (not necessarily random) of instances from a larger set? If the dataset is a sample, then what is the larger set? Is the sample representative of the larger set (e.g., geographic coverage)? If so, please describe how this representativeness was validated/verified. If it is not representative of the larger set, please describe why not (e.g., to cover a more diverse range of instances, because instances were withheld or unavailable).} It is collected from an online application in a week without sampling. It is representative.

\paragraph{Are relationships between individual instances made explicitly (e.g., users' movie ratings, social network links)? If so, please describe how these relationships are made explicit.} N/A.

\paragraph{Are there recommended data splits (e.g., training, development/validation, testing)? If so, please provide a description of these splits, explaining the rationale behind them.} The code of data splits is provided in 
\url{https://github.com/fuxiAIlab/RL4RS/blob/main/reproductions/run_split.sh}.

\paragraph{Are there any errors, sources of noise, or redundancies in the dataset? If so, please provide a description.} No.

\paragraph{Is the dataset self-contained, or does it link to or otherwise rely on external resources (e.g., websites, tweets, other datasets)? If it links to or relies on external resources, a) are there guarantees that they will exist, and remain constant, over time; b) are there official archival versions of the complete dataset (i.e., including the external resources as they existed at the time the dataset was created); c) are there any restrictions] (e.g., licenses, fees) associated with any of the external resources that might apply to a future user? Please provide descriptions of all external resources and any restrictions associated with them, as well as links or other access points, as appropriate.} RL4RS is self-contained.

\paragraph{Does the dataset contain data that might be considered confidential (e.g., data that is protected by legal privilege or by doctorpatient confidentiality, data that includes the content of individuals' non-public communications)? If so, please provide a description.} No, all the samples are public available.

\paragraph{Does the dataset contain data that, if viewed directly, might be offensive, insulting, threatening, or might otherwise cause anxiety? If so, please describe why.} No.

\paragraph{Does the dataset relate to people? If not, you may skip the remaining questions in this section.} No.

\subsection{Uses}

\paragraph{Has the dataset been used for any tasks already? If so, please provide a description?} It is proposed to use for reinforcement learning based recommendation task.

\paragraph{Is there a repository that links to any or all papers or systems that use the dataset? If so, please provide a link or other access point.} It is a new dataset. A older and smaller dataset collected from the same application is used in a kaggle competition \url{https://www.kaggle.com/c/bigdata2021-rl-recsys/overview}.

\paragraph{What (other) tasks could the dataset be used for?} Many other tasks like neural combinatorial optimization can be also used.

\paragraph{Is there anything about the composition of the dataset or the way it was collected and preprocessed/cleaned/labeled that might impact future uses? For example, is there anything that a future user might need to know to avoid uses that could result in unfair treatment of individuals or groups (e.g., stereotyping, quality of service issues) or other undesirable harms (e.g., financial harms, legal risks) If so, please provide a description. Is there anything a future user could do to mitigate these undesirable harms?} N/A

\paragraph{Are there tasks for which the dataset should not be used? If so, please provide a description.} N/A

\subsection{Distribution}

\paragraph{Will the dataset be distributed to third parties outside of the entity (e.g., company, institution, organization) on behalf of which the dataset was created?} All datasets are released to the public.

\paragraph{How will the dataset will be distributed (e.g., tarball on website, API, GitHub)?} All datasets are released on Zenodo \url{https://zenodo.org/record/6622390#.YqBBpRNBxQK} and  our github repository  \url{https://github.com/fuxiAIlab/RL4RS}. The dataset DOI is 10.5281/zenodo.6622390.

\paragraph{When will the dataset be distributed?} It has been released now.

\paragraph{Will the dataset be distributed under a copyright or other intellectual property (IP) license, and/or under applicable terms of use (ToU)?} Our dataset is distributed under the CC BY-SA 4.0 license, see  \url{https://github.com/fuxiAIlab/RL4RS/blob/main/LICENSE}.

\paragraph{Have any third parties imposed IP-based or other restrictions on the data associated with the instances? If so, please describe these restrictions, and provide a link or other access point to, or otherwise reproduce, any relevant licensing terms, as well as any fees associated with these restrictions.} No.

\paragraph{Do any export controls or other regulatory restrictions apply to the dataset or to individual instances? If so, please describe these restrictions, and provide a link or other access point to, or otherwise reproduce, any supporting documentation.} No.

\subsection{Maintenance}

\paragraph{How can the owner/curator/manager of the dataset be contacted (e.g., email address)?}

Kai Wang (\texttt{wangkai02@corp.netease.com}) and Runze Wu (\texttt{wurunze1@corp.netease.com}) will be responsible for maintenance.

\paragraph{Will the dataset be updated (e.g., to correct labeling errors, add new instances, delete instances)?}
Yes. If we include more tasks or find any errors, we will correct the dataset and update the leaderboard accordingly. It will be updated on our website.

\paragraph{If others want to extend/augment/build on/contribute to the dataset, is there a mechanism for them to do so?}
They can contact us via email for the contribution.

\section{Working with RL-based RS Data Using RL4RS}
\label{appendix_c}

In order to facilitate the entry of new practitioners to the field of RL-based recommender system, we provide some detailed guidelines for working with the datasets we provide and for managing new ones in the framework of the \texttt{rl4rs} package. We have provided some tools required for RL-based RS task (e.g., data understanding tools), and will be continuing to develop and support more functionalities.

\subsection{Assembling New Datasets}

\paragraph{Data Characteristic Examine.}
Recommender systems are an area that relies heavily on data. When appling RL into RS, one big challenge is problem formulation. It is easy to accidentally prepare data that does not conform well to the MDP definition, and applying RL on ill-formulated problems isa costly process. Here, we develop a data understanding tool. Using a data-driven method together with heuristics, the tool checks whether the properties of RS datasets conform to the RLframework. RL4RS provides three usecases for RS datasets to help users quickly use the tool to check their new datasets. In practice, it may help us understand the dataset in early stages, catch improperly defined RL problems, and lead to a better prob-lem definition (e.g., reward definition)

\paragraph{Batched Environment Preparation.}
Different from gym environments, simulation environments of RL-based RS tasks are often based on a pre-trained neural network. In most recommended scenarios, this often includes several thousand dimensional features, millions of samples, several GB files, and tens of milliseconds of inference delay. These characteristics make this environment difficult to be accelerated by the existing RL libraries, and lead to the model can not be trained in an acceptable time. Based on the popular RLLib library, RL4RS developes two solutions: file-based RL environment and HTTP-based vector environment. The former enables random sampling and sequential access to datasets exceeding memory size, and can be easy to extend to distributed file systems. The latter allows the environment processes batch data at one time and to be deployed on multiple servers to accelerate the generation of samples. Users only need to process new dataset into the specified data format to use these two functionalities.

\subsection{Developing and Benchmarking New Algorithms}

\paragraph{Comparing Algorithms.} The best practices for RL-based RS are still poorly explored. Not only do we compare different RL models on a given environment model, but we also need to compare different environment construction methods, different batch RL methods which learn directly from offline data, and even different MDP definitions. Just for comparing RL, we need to consider whether actions are defined as continuous or discrete, use hidden layer or raw feature as observation, use action mask as hard or soft constraint, etc. RL4RS suits provide all option combinations to help users to explore the best practices for their new dataset.

\paragraph{Evaluation Outside the Environment.}
Predictions can be tested with various  metrics. Current RL evaluation methodology, i.e., training and evaluating on the same environment, is not yet well-suited for RL-based RS tasks. RL4RS provides some other different evaluation methods, including evaluation of simulation environments, offline policy evaluation and evaluating on the environment built from test set.
In the future, we will add more evaluation metrics. To calculate new metrics for older models, when benchmarking an algorithm, specific predictions should be stored and not only metrics. In RL4RS, we provide standardized evaluation classes to calculate the appropriate metrics from the outputs of a given algorithm. Users only need to implement their custom environment class for the new dataset by inheriting the base class.

\section{Experimental Details}
\label{appendix_d}

\subsection{Implementation Details of Data Preprocess}
Given the raw logged data described in Section \ref{sec3}, 
specifically, we transforms the logged data collected in the following row format:

\begin{itemize}
    \item \textbf{MDP ID}: A unique ID for the Markov Decision Process (MDP) chain that this training example is a part of.
    
    \item \textbf{Sequence ID}: A number representing the location of the state in the MDP (i.e. a timestamp).
    
    \item \textbf{State}: The features of current step that consists of user features and context features.
    
    \item \textbf{Observation}: The 256-dim hidden layer embedding of the pre-trained environment model with user features and context features of current step as input. Or we can use the raw state features as observation. 
    
    \item \textbf{Action}: The item ID taken at the current step. Or we can use item embeddings can be used to represent item for continuous environment setting.
    
    \item \textbf{Action Features}: The features of item taken at the current step, including item embedding. 
    
    \item \textbf{Action Probability}: The probability that the behavior policy took the action. 
    
    \item \textbf{Action Mask}: An list of possible items at the current step.
    
    \item \textbf{Reward}: The reward of the current step.
    
    \item \textbf{Next State}: The state features after acting the logged action.
    
    \item \textbf{Next Observation}: The 256-dim hidden layer embedding of next state features.
    
    \item \textbf{Next Action}: The item ID recommended at the next step.
    
    \item \textbf{Next Action Features}: The features of item recommended at the next step.
    
    \item \textbf{Next Action Probability}: The probability of the item that is recommended at the next step.
    
    \item \textbf{Next Action Mask}: A list of items that are allowed to recommend at the next step.
    
    \item \textbf{Terminal}: A 0/1 number representing whether it is the last state.
    
\end{itemize}
The details of data splits and the above data transformations are provided in the open-source code.
\subsection{Implementation Details of Simulation Environment Construction}
\label{appendix:imple}
The detailed network architectures  of baselines are provided in \url{https://github.com/fuxiAIlab/RL4RS/tree/main/rl4rs/nets}.
Here we introduce the three tasks for simulation environment construction.

\paragraph{Slate-wise Classification Task.}  The question can be transformed into a multi-class classification task, where the feedbacks of users to the nine items in the page are considered as a whole. Remove impossible situations, there are 22 possible classes. Given the nine items in the page \(\left<a_{1}, a_{2}, \ldots, a_{9}\right>\), corresponding user feedbacks \(\left<o_{1}, o_{2}, \ldots, o_{9}\right>\), and the model 0/1 prediction \(\left<\hat{o}_{1}, \hat{o}_{2}, \ldots, \hat{o}_{9}\right>\), the accuracy metric is defined as the identical accuracy between  \(\left<o_{1}, o_{2}, \ldots, o_{9}\right>\) and \(\left<\hat{o}_{1}, \hat{o}_{2}, \ldots, \hat{o}_{9}\right>\).

\paragraph{Item-wise Rank Task.} The question can be transformed into a item ranking task (or a multi-label task), where the feedbacks of users to the nine items in the page are treated as different labels. Given the nine items in the page \(\left<a_{1}, a_{2}, \ldots, a_{9}\right>\), corresponding user feedbacks \(\left<o_{1}, o_{2}, \ldots, o_{9}\right>\), and the user's purchase probability prediction of the model \(\left<\hat{y}_{1}, \hat{y}_{2}, \ldots, \hat{y}_{9}\right>\), the AUC, precision and recall metrics can be calculated.

\paragraph{Item-wise Classification Task.}  The question can be transformed into a binary classification task, where the feedbacks of users to the nine items in the page are considered item by item. Given the nine items in the page as recommendation context, the model outputs the user's purchase probability $\hat{y}_{j}$ for item $j$. The AUC and accuracy metrics are calculated based on the purchase probability $y$ and ground-truth label $o$.

\subsection{Fine-tuning Details of Model-free Reinforcement Learning}
In RL4RS, model-free RL algorithms are implemented based on the popular open-source library RLLib~\cite{liang2018rllib}. The gamma is set to 1. The fine-tuned models and reproduction scripts are all included in the Github repository. All baselines share the same encoder network architecture. The detailed parameters are shown in Table \ref{tab:fine_tune_modelfree}.

\begin{table}[]
\scriptsize
\centering
\caption{The fine-tuning details of model-free reinforcement learning algorithms under the setting of discrete action space.}
\label{tab:fine_tune_modelfree}
\begin{tabular}{lccc}
\toprule
  & Exploration & Training Config & Model Config \\ \hline
PG & SoftQ & \begin{tabular}[c]{@{}l@{}}steps:6400000\\ batch\_size:256 \\ hidden\_units:128 \\ learning\_rate: 0.0004 \end{tabular} & default \\
\midrule
DQN & SoftQ & \begin{tabular}[c]{@{}l@{}}steps:6400000\\ batch\_size:256 \\ hidden\_units:128 \\ learning\_rate: 0.0001 \end{tabular} & \begin{tabular}[c]{@{}l@{}}target\_update\_freq:200\\ double\_q:True \\ dueling:False \\ buffer\_size: 100000 \\ n\_step: 1 \end{tabular}  \\
\midrule
RAINBOW & SoftQ & \begin{tabular}[c]{@{}l@{}}steps:6400000\\ batch\_size:256 \\ learning\_rate: 0.0001 \\ no custom encoder \end{tabular} & \begin{tabular}[c]{@{}l@{}}target\_update\_freq:200\\ double\_q:True \\ dueling:True \\ buffer\_size: 100000 \\ n\_step: 3 \\ num\_atoms:8 \end{tabular}  \\
\midrule
A2C & SoftQ & \begin{tabular}[c]{@{}l@{}}steps:6400000\\ batch\_size:256 \\ hidden\_units:128 \\ learning\_rate: 0.0001 \end{tabular} & \begin{tabular}[c]{@{}l@{}}use\_critic:True \\ use\_gae:True \\ vf\_loss\_coeff:0.5 \\ entropy\_coeff:0.01\quad\quad\quad \\ grad\_clip:10 \end{tabular} \\
\midrule
PPO & SoftQ & \begin{tabular}[c]{@{}l@{}}steps:6400000\\ batch\_size:256 \\ hidden\_units:128 \\ learning\_rate: 0.0001 \end{tabular} & \begin{tabular}[c]{@{}l@{}}use\_critic:True\\ use\_gae:True \\ kl\_coeff:0.2 \\ vf\_loss\_coeff:0.5 \\ clip\_param:0.3 \\ kl\_target:0.1 \\ sgd\_minibatch\_size:256 \\ num\_sgd\_iter:1 \end{tabular}  \\
\midrule
Exact-k & EpsilonGreedy & \begin{tabular}[c]{@{}l@{}}steps:5120000\\ batch\_size:512 \\ hidden\_units:256 \\ actor\_lr: 0.001 \\ critic\_lr: 0.005 \quad\quad\quad \end{tabular} & \begin{tabular}[c]{@{}l@{}}num\_heads:4\\ num\_layers:1 \\ num\_blocks:2 \\ use\_mha:True\end{tabular} \quad\quad\quad\quad\quad \\
\bottomrule
\end{tabular}
\vspace{-4mm}
\end{table}

\begin{table}[]
\scriptsize
\centering
\caption{The fine-tuning details of model-free reinforcement learning algorithms under the setting of continuous action space.}
\label{tab:fine_tune_modelfree}
\begin{tabular}{lccc}
\toprule
  & Exploration & Training Config & Model Config \\ \hline
PG-conti & StochasticSampling & \begin{tabular}[c]{@{}l@{}}steps:6400000\\ batch\_size:256 \\ hidden\_units:128 \\ learning\_rate: 0.0004 \\ action\_emb\_size:32 \end{tabular} & default  \\
\midrule
A2C-conti & StochasticSampling & \begin{tabular}[c]{@{}l@{}}steps:6400000\\ batch\_size:256 \\ hidden\_units:128 \\ learning\_rate: 0.0001 \\ action\_emb\_size:32 \end{tabular} & \begin{tabular}[c]{@{}l@{}}use\_critic:True \\ use\_gae:True \\ vf\_loss\_coeff:0.5 \\ entropy\_coeff:0.01 \quad\quad\quad \\ grad\_clip:10 \end{tabular} \\
\midrule
PPO-conti & StochasticSampling & \begin{tabular}[c]{@{}l@{}}steps:6400000\\ batch\_size:256 \\ hidden\_units:128 \\ learning\_rate: 0.0001 \\ action\_emb\_size:32 \end{tabular} & \begin{tabular}[c]{@{}l@{}}use\_critic:True\\ use\_gae:True \\ kl\_coeff:0.2 \\ vf\_loss\_coeff:0.5 \\ clip\_param:0.3 \\ kl\_target:0.1 \\ sgd\_minibatch\_size:256 \\ num\_sgd\_iter:1 \end{tabular}  \\
\midrule
DDPG & OrnsteinUhlenbeckNoise & \begin{tabular}[c]{@{}l@{}}steps:6400000\\ batch\_size:256 \\ hidden\_units:128 \\ learning\_rate: 0.001 \\ action\_emb\_size:32 \end{tabular} & \begin{tabular}[c]{@{}l@{}} twin\_q:False\\ target\_noise:0.2 \quad\quad\quad\quad \end{tabular} \\
\midrule
TD3 & OrnsteinUhlenbeckNoise & \begin{tabular}[c]{@{}l@{}}steps:6400000\\ batch\_size:256 \\ hidden\_units:128 \\ learning\_rate: 0.001 \\ action\_emb\_size:32 \end{tabular} & \begin{tabular}[c]{@{}l@{}} \quad random\_timesteps:10000 \end{tabular} \\
\bottomrule
\end{tabular}
\vspace{-4mm}
\end{table}

\subsection{Fine-tuning Details of Batch Reinforcement Learning}
In RL4RS, batch RL algorithms are implemented based on the open-source batch RL library d3rlpy~\cite{seno2020d3rlpy}. The gamma is set to 1. The fine-tuned models and reproduction scripts are all included in the Github repository. All baselines share the same encoder network architecture. The detailed parameters are shown in Table \ref{tab:fine_tune_batch}.
\begin{table}[h]
\centering
\small
\caption{The fine-tuning details of batch reinforcement learning algorithms.}
\label{tab:fine_tune_batch}
\begin{tabular}{lcc}
\toprule
   & Training Config & Model Config \\ \hline
BC & \begin{tabular}[c]{@{}l@{}}epoch:10\\ batch\_size:256 \\ hidden\_units:256 \\ learning\_rate: 0.0001 \end{tabular} & \begin{tabular}[c]{@{}l@{}} bc\_beta:0 \\ reward\_scaler:`None'\quad\quad\quad\quad \end{tabular} \\
\midrule
BCQ & \begin{tabular}[c]{@{}l@{}}epoch:10\\ batch\_size:256 \\ hidden\_units:256 \\ learning\_rate: 6.25e-5 \end{tabular} & \begin{tabular}[c]{@{}l@{}}action\_flexibility:0.3 \\ bcq\_beta:0.5 \\ share\_encoder:False \\ reward\_scaler:`None' \\ target\_update\_interval:8000\end{tabular}  \\
\midrule
CQL & \begin{tabular}[c]{@{}l@{}}epoch:10\\ batch\_size:256 \\ hidden\_units:256 \\ learning\_rate: 6.25e-5 \end{tabular} & \begin{tabular}[c]{@{}l@{}}gamma:1.0 \\ share\_encoder:True \\ reward\_scaler:`standard'\quad\quad \\ cql\_alpha:1 \end{tabular} \\
\midrule
MOPO & \begin{tabular}[c]{@{}l@{}}epoch:10\\ batch\_size:256 \\ hidden\_units:256 \\ learning\_rate: 3e-4 \quad  \end{tabular} & \begin{tabular}[c]{@{}l@{}} n\_critics:2 \\ rollout\_horizon:5 \\ obs\_scaler:`min\_max' \\ reward\_scaler:`standard'  \quad \end{tabular}  \\
\midrule
COMBO & \begin{tabular}[c]{@{}l@{}}epoch:10\\ batch\_size:256 \\ hidden\_units:256 \\ learning\_rate: 3e-4 \quad  \end{tabular} & \begin{tabular}[c]{@{}l@{}}conservative\_weight:1.0 \\ n\_critics:2 \\  obs\_scaler:`min\_max' \\ reward\_scaler:`standard' \quad  \\ n\_action\_samples:10\end{tabular}  \\
\bottomrule
\end{tabular}
\vspace{-4mm}
\end{table}

\section{Data Understanding}
\label{appendix_f}
One big challenge of applied RL is problem formulation. Traditional recommendation scenarios and datasets are not always suitable for modeling as MDP problems where some sort of long-term reward is optimized in a sequential setting.
It is easy to accidentally prepare data that does not conform well to the MDP definition, and applying RL on ill-formulated problems is a costly process.
Here, we develop a data understanding tool. Using a data-driven method together with heuristics, the tool checks whether the properties of RS datasets conform to the RL framework.

\subsection{Long-term Impact}

An optimal policy should take into account  the long-term impact of a recommendation on the user's future choices.
To maximize the accumulative rewards, we might suggest an item whose immediate reward is low but leads to more likely or more profitable rewards in the future.
For example, when the products sold are books, by recommending a book for which there is a sequel, we may increase the likelihood that this sequel will be purchased later.
Heuristically, the way to measure whether a recommendation problem should be modeled as an RL problem is to see whether a recommendation decision has a long-term impact. 
In terms of reinforcement learning formula, the recommendation of step $t$ is to maximize $r(s_{t},a_{t})+V^{*}(s_{t+1})$, where $r(s_{t},a_{t})$ is the expected reward when recommending item $a_{t}$ at state $s_{t}$, and $V^{*}(s_{t+1})$ represents the maximum reward of next state under current policy.
Since there is always a large state value to represent users retention, we further denote $A(s_{t},a_{t})=V^{*}(s_{t+1})-V^{*}(s_{t}, \cdot)$ as the difference (advantage) between the future reward of performing $a_{t}$ and the averaged future rewards, i.e., the long-term impact of performing $a_{t}$ at state $s_{t}$.
When there is no long-term impact or the long-term impact is small, the RL problem degenerates into an ordinary sequential recommendation problem, which only maximizes the reward of the current step.

Though we can estimate the long-term impact by training an RL model, it is too complex. Instead, we establish a data understanding tool to quantify the long-term impact without the requirement of establishing environment models or learning a value function. This tool is easy to use by simplifying RL as sequence modeling problems and expected rewards as decoded sequence scores. 
Similar ideas have been developed in Trajectory Transformer~\cite{transformer-rl} and Decision Transformer~\cite{transformer-decision}, in which states, actions, and returns are fed into a GPT~\cite{Radford2018ImprovingLU} architecture and actions are decoded autoregressively.

First, the tool fits a Transformer-based sequence-to-sequence model~\cite{transformer} on each offline RS dataset.
It encodes the user context features and decodes K items that users may click or buy, which means that the recommendation system will recommend these K items in turn in the next K steps (considering that most RS datasets do not provide a complete page context, here, only one item is recommended at a time, eliminating the slate effect between items within one page). 
We consider using the decode sequence generated by greedy search to represent greedy recommendation (SL-based strategy), and the sequence generated by beam search to represent the recommendation result generated by the optimal RL policy.
We use beam-search width of 100. 
When there is a significant long-term item impact in the dataset, the items recommended in the previous steps may not have high immediate impacts, but their long-term impacts are large. 
That is, the immediate reward of the first k item in the best decode sequence may account for a small proportion in the final score of the sequence (experiment \uppercase\expandafter{\romannumeral1}).
On the other hand, we can compare the sequences score of greedy strategy (even the sequence composed of only hot items which have high immediate reward) with the score of optimal sequence to check whether the greedy strategy is good enough (experiment \uppercase\expandafter{\romannumeral2}).

\begin{table}[tb]
\centering
\caption{The comparison of item's long-term impact between different datasets.}
\label{tab:mdp}
\begin{tabular}{lccc}
\toprule
Score. of & RecSys15 & MovieLens & RL4RS-Slate \\ \hline
1-Pearson & 0.11 & 0.13 & 0.03   \\
1-Spearman & 0.10 & 0.11 & 0.02    \\
2-Pearson & 0.16 & 0.17 & 0.06   \\
2-Spearman & 0.15 & 0.16 & 0.03   \\
3-Pearson & 0.24 & 0.25 & 0.14   \\
3-Spearman & 0.23 & 0.23 & 0.08   \\
4-Pearson & 0.39 & 0.38 & 0.36   \\
4-Spearman & 0.39 & 0.36 & 0.33   \\
5-Pearson & 1.00 & 1.00 & 1.00   \\
5-Spearman & 1.00 & 1.00 & 1.00   \\
\bottomrule
\end{tabular}
\label{table:mdp1}
\vspace{-4mm}
\end{table}

\begin{table}[tb]
\centering
\caption{The comparison of decode sequence scores between different datasets.}
\label{tab:mdp}
\begin{tabular}{lccccc}
\toprule
Score. of & 5\% & 20\% & greedy. & hot 5\% & hot 20\%\\ \hline
RecSys15 & 1.00 & 0.53 & 1.26  & 0.81 & 0.42 \\
MovieLens & 1.00 & 0.64 & 0.99  & 0.98 & 0.62  \\
Last.fm & 1.00 & 0.47 & 1.09  & 0.90 & 0.42  \\
CIKMCup2016 & 1.00 & 0.30 & 4.50  & 0.99 & 0.28  \\
RL4RS-Slate & 1.00 & 0.76 & 0.62 & 0.01 & 0.01  \\
\bottomrule
\end{tabular}
\label{table:mdp21}
\vspace{-4mm}
\end{table}

\subsection{Experiment}
We build experiments on the following datasets, MovieLens-25m, RecSys15 YooChoose, and RL4RS-Slate (has the same result as RL4RS-SeqSlate under this experiment setting).
Without losing generality, we only consider the long-term impact within 5 steps. For each dataset, we choose 10000 users randomly as test set, and others as train set. For each user, we calculate the greedy search result and the top 100 item sequences (beam search width as 100).  In the first experiment, we report the Pearson Correlation Coefficient and Spearman Rank Correlation Coefficient between the scores of the first k items (immediate reward) and the final sequence scores, denoted as k-Pearson and k-Spearman respectively. If the item's long-term impact is significant, the coefficients should be small, i.e., the immediate reward of the first item in the decode sequence accounts for a small proportion of the final score of the sequence. We report the results on the top 100 item sequences generated by beam search, as shown in Table~\ref{table:mdp1}. 
It can be seen that RecSys15 and MovieLens show positive correlations at the beginning. RL4RS dataset achieves a significantly lower spearman rank coefficient at the first one and two items than the other two datasets. It means, in RL4RS dataset, the first k items recommended do not have high immediate impacts, but have great long-term impacts. Although the first experiment can not absolutely indicate whether a dataset is suitable for reinforcement learning, it provides a tool for comparison of the degree of long-term impact characteristics between different datasets.

In the second experiment, we aim to distinguish the dataset property by checking whether the greedy strategy is already good enough. We report the averaged score of the top 5 (5\% quantile) decode sequence, the top 20 (20\% quantile) sequence, and the score of sequence generated by greedy strategy. We also calculate the averaged score of the first 5\% quantile and 20\% quantile sequences when limiting the candidate items to hot items (100 items with the highest probability at the first decode step). All these scores are normalized by the average score of the top 5 sequences for each dataset.
The results are shown in Table~\ref{table:mdp21}. It can be seen that there is a significant gap in the scores between the best 5\% item sequences and greedy strategy for RL4RS dataset, let alone the sequences composed of hot items. 
It shows that it is necessary to model the RL4RS dataset as an RL problem. For traditional RS datasets, on the contrary, the score of greedy strategy is very close to that of the best RL policy (top 5 decode sequence).
It indicates that the greedy strategy (i.e., sequential RS methods) is well enough to model these scenarios.

Although the data understanding tool is based on heuristics, in practice, it may help us understand the dataset in the early stages, catch improperly defined RL problems, and lead to a better problem definition (e.g., reward definition).

\section{Policy Evaluation Framework}
\label{appendix_g}
In the previous section, we provide a simple performance comparison between model-free RL and batch RL. These results are collected following the traditional RL evaluation framework, that is, training and evaluating on the same environment. However, in real applications, it is usually rare to have access to a perfect simulator. A policy should be well evaluated before deployment because policy deployment affects the real world and can be costly.
In applied settings, policy evaluation is not a simple task that can be quantified with only a single indicator (e.g., reward).
As shown in Figure~\ref{Fig:eval}, we attempt to propose a new evaluation framework for RL-based RS, including environment simulation evaluation, counterfactual policy evaluation (offline policy evaluation), and evaluation on simulation environments built from test set (online policy evaluation).

\begin{figure}[h]
  \includegraphics[scale=0.16]{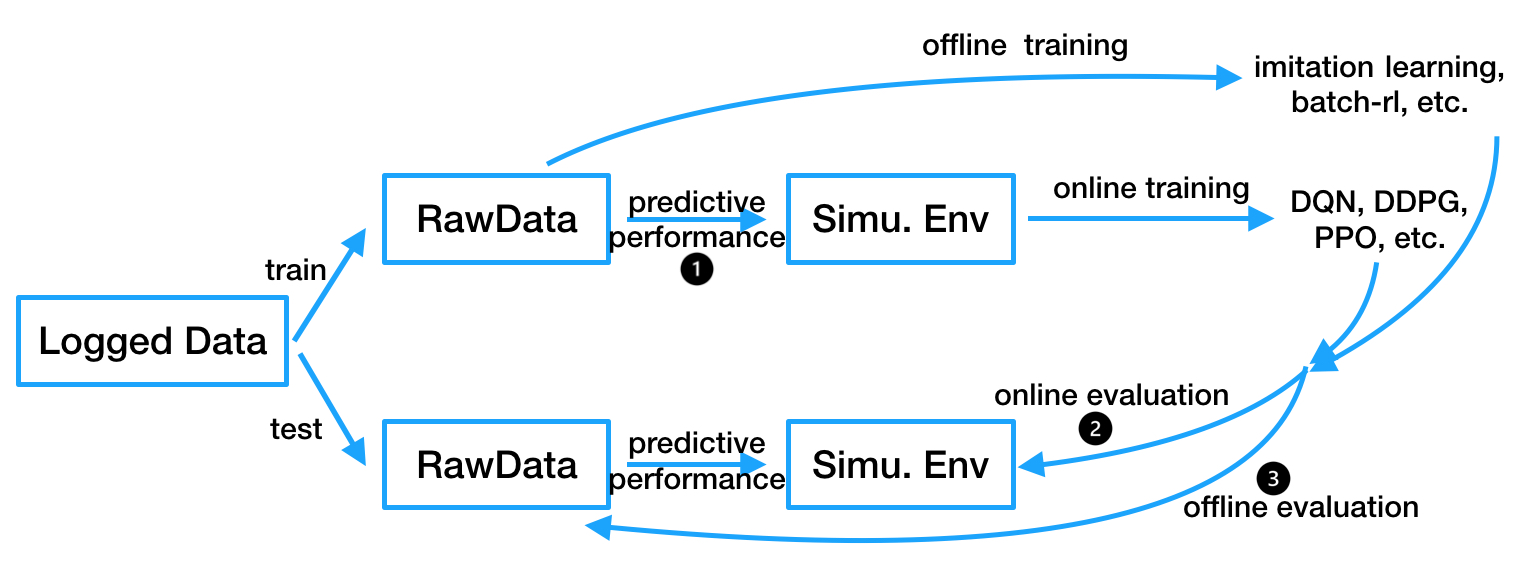}
  \caption{The evaluation framework for the RL-based RS task. \label{Fig:eval}}
\end{figure}

\subsection{Evaluation On Another Simulation Environment}
A possible evaluation method is to establish an environment on the test set to evaluate the algorithm. Different from testing in the same environment, we emphasize the establishment of environment model on test set according to the evaluation framework of supervised learning. In many recommendation scenarios, the training set and test set need to be divided according to time, which makes this more important. Another popular evaluation method adopted by most current works is to divide users before policy training. Although the training and test data are not overlapped, they are sharing the same environment. This method still exists a risk of over-fitting, because the estimated reward of the test set will be affected by the train set through the shared environment.

According to the method described in Section 5.1, we construct the environment model on the training set and the test set  respectively. We use the most strict dataset dividing setting, that is, Dataset-SL as the train set and Dataset-RL as the test set. Not only they are not overlapped on time, but also there are considerable differences in data distribution between them. For model-free RL and batch RL, we compare the performance of the following three settings: (1) constructing environment model on Dataset-SL plus Dataset-RL, training the policy on Dataset-SL, evaluating on Dataset-SL; (2) constructing environment model on Dataset-SL, training the policy on Dataset-SL, evaluating on Dataset-RL;
(3) constructing environment model on Dataset-SL, training the policy on Dataset-SL, evaluating on the environment model built from Dataset-RL. For environment simulation, we provide the result of the following two settings: (1) constructing environment model on Dataset-SL, and then evaluating on Dataset-RL; and (2) constructing environment model on Dataset-RL, and then evaluating on Dataset-SL; The results of training on all dataset and evaluate on Dataset-SL and Dataset-RL are also provided as a baseline.
The results are as follows.

\begin{table}[tb]
\centering
\caption{Reward estimation performance comparison of DIEN-based environment simulator between different environment simulation settings (setting 1, 2, 3, and 4 represent the setting of training on all dataset then testing on Dataset-SL, training on all dataset then testing on Dataset-RL,  training on Dataset-RL then testing on Dataset-SL, and training on Dataset-SL then testing on Dataset-RL respectively). On-going experiments.}
\begin{tabular}{|c|clc|clc|}
\hline
\multirow{2}{*}{} & \multicolumn{3}{c|}{RL4RS-Slate}                                      & \multicolumn{3}{c|}{RL4RS-SeqSlate}                                      \\ \cline{2-7} 
                  & \multicolumn{1}{c|}{mean.} & \multicolumn{1}{c|}{abs.} & std. & \multicolumn{1}{c|}{mean} & \multicolumn{1}{c|}{abs.} & std. \\ \hline
Setting 1               & \multicolumn{1}{c|}{-2.3}    & \multicolumn{1}{c|}{38.1}         & 66.5    & \multicolumn{1}{c|}{-13.3}    & \multicolumn{1}{c|}{63.5}         & 99.3    \\ \hline
Setting 2        & \multicolumn{1}{c|}{-2.7}    & \multicolumn{1}{c|}{36.3}         & 61.4    & \multicolumn{1}{c|}{-6.8}    & \multicolumn{1}{c|}{65.0}         & 97.7    \\ \hline
Setting 3              & \multicolumn{1}{c|}{1.2}    & \multicolumn{1}{c|}{42.4}         & 69.4    & \multicolumn{1}{c|}{2.4}    & \multicolumn{1}{c|}{71.5}         & 106.5    \\ \hline
Setting 4           & \multicolumn{1}{c|}{4.5}    & \multicolumn{1}{c|}{40.6}         & 64.4    & \multicolumn{1}{c|}{23.3}    & \multicolumn{1}{c|}{81.0}         & 106.9   \\ \hline
\end{tabular}
\label{table:simulator_eval_2}
\vspace{-4mm}
\end{table}

From Table~\ref{table:simulator_eval_2}, it can be found that comparing setting 3/4 and setting 1/2, there is basically a 10\% increase in the absolute mean indicator and a 5\% increase in the standard deviation indicator on RL4RS-Slate dataset. On RL4RS-SeqSlate  dataset, the indicator difference widens further, especially with setting 4 versus setting 2. Since the initial error of environment construction is already large, the impact of dataset division is difficult to measure directly. It needs to be further revealed by the results of the RL algorithms.

Table~\ref{table:offline_training_2} shows the results of the three batch RL algorithms under the three experimental settings mentioned above and the baseline experimental settings (setting 1). Because the experiments are time-consuming, only a few rounds of experiments have been done at present. We will provide the averaged results of more rounds in the future. From the comparison between setting 3/4 and setting 2, the choice of environment construction dataset has a great impact on the final result. Compared with setting 2 and setting 1, the results do not show much difference (except CQL algorithm). It may be because they both use all dataset to construct the environment model. Although setting 2 only trains on Dataset-SL, it also learns a lot from Dataset-RL.

\begin{table}[h]
\centering
\caption{Performance comparison between batch RL algorithms on different experiment settings (setting 1, 2, 3, and 4 represent the setting of building environment and training on all datasets then test on Dataset-RL, building environment on all datasets and training on Dataset-SL then test on Dataset-RL, building environment and training on Dataset-SL then test on Dataset-RL, and building environment and training on Dataset-SL then test on the environment built from Dataset-RL respectively). On-going experiments.}
\begin{tabular}{|c|clc|clc|}
\hline
\multirow{2}{*}{} & \multicolumn{3}{c|}{RL4RS-Slate}                                      & \multicolumn{3}{c|}{RL4RS-SeqSlate}                                      \\ \cline{2-7} 
                  & \multicolumn{1}{c|}{BC} & \multicolumn{1}{c|}{BCQ} & CQL & \multicolumn{1}{c|}{BC} & \multicolumn{1}{c|}{BCQ} & CQL \\ \hline
Setting 1               & \multicolumn{1}{c|}{98.0}    & \multicolumn{1}{c|}{132.9}         & 107.2    & \multicolumn{1}{c|}{240.7}    & \multicolumn{1}{c|}{277.2}         & 259.4    \\ \hline
Setting 2        & \multicolumn{1}{c|}{98.8}    & \multicolumn{1}{c|}{133.4}         & 98.8    & \multicolumn{1}{c|}{244.7}    & \multicolumn{1}{c|}{273.7}         & 268.3    \\ \hline
Setting 3           & \multicolumn{1}{c|}{103.7}    & \multicolumn{1}{c|}{200.7}         & 109.3    & \multicolumn{1}{c|}{230.6}    & \multicolumn{1}{c|}{288.9}         & 274.2    \\ \hline
Setting 4              & \multicolumn{1}{c|}{107.1}    & \multicolumn{1}{c|}{133.6}         & 108.6    & \multicolumn{1}{c|}{139.5}    & \multicolumn{1}{c|}{196.6}         & 174.6    \\ \hline
\end{tabular}
\label{table:offline_training_2}
\vspace{-4mm}
\end{table}

\end{document}